\documentclass[a4paper]{article}
\pdfoutput=1 % if your are submitting a pdflatex (i.e. if you have
\usepackage{custom_style} 
\usepackage[utf8]{inputenc}

\usepackage{xcolor}
\usepackage{amsmath}
\usepackage{amssymb}
\usepackage{bm}

\allowdisplaybreaks   % allows align equations to split over pages

\usepackage{relsize}
\usepackage{graphicx}
\DeclareMathOperator*{\argmax}{arg\,max} % thin space, limits underneath in
\def\bi#1{\hbox{\boldmath{$#1$}}}

\usepackage{enumitem}

\title{%Maximum Posterior Significance: 
The look-elsewhere effect from a unified Bayesian and frequentist perspective}

\author[a]{Adrian E.~Bayer}
\author[a,b]{and Uro\v{s} Seljak}

\affiliation[a]{Berkeley Center for Cosmological Physics, 
Department of Physics, University of California, Berkeley,\\ Berkeley, CA 94720, USA}
\affiliation[b]{Physics Division, Lawrence Berkeley National Laboratory,  1 Cyclotron Road,\\ Berkeley, CA 94720, USA}

\emailAdd{abayer@berkeley.edu}
\emailAdd{useljak@berkeley.edu}

\abstract{
When searching over a large parameter space
for anomalies such as events, peaks, objects, or particles, there is a large probability that spurious signals with seemingly high significance will be found. This is known as the look-elsewhere effect and is prevalent throughout cosmology, (astro)particle physics, and beyond. To avoid making false claims of detection, one must account for this effect when assigning the statistical significance of an anomaly.
This is typically accomplished by considering the 
trials factor, which is generally computed numerically via potentially expensive simulations. In this paper we develop 
a continuous generalization 
of the Bonferroni and \v{S}id\'{a}k 
corrections  
by applying the Laplace approximation to evaluate the Bayes factor,
and in turn relating the trials factor to the prior-to-posterior volume ratio.
We use this to define a test statistic whose frequentist properties have a simple interpretation in terms of the 
global $p$-value, or statistical significance. 
We apply this method to various physics-based 
examples and show it to work well for the full range of $p$-values, i.e.~in both the asymptotic and non-asymptotic regimes. 
We also show that this method naturally accounts for other model complexities such as additional degrees of freedom, generalizing Wilks' theorem.
This provides a fast way to quantify statistical significance in light of the look-elsewhere effect, without resorting to expensive simulations.
}

\begin{document}
\maketitle

\flushbottom

\section{Introduction}

A common problem in statistical analysis is to find evidence for
a physical signal
%events, peaks, features, resonances etc., 
in a large, continuous parameter space, where the true position of the signal is not known a priori. 
By searching over a wide 
%and continuous 
parameter 
%phase
space one increases the probability of finding large signals caused by random statistical fluctuations, as opposed to a physical source. This is known as the look-elsewhere effect -- or sometimes the ``problem of multiple comparisons'' in discrete cases -- 
and must be accounted for when performing a hypothesis test 
\cite{Miller_1981, mult_hyp_test}. Ignoring this effect would lead to an overestimation of the statistical significance, sometimes by a considerable amount, and thus incorrectly concluding the detection of a physical signal. 

The look-elsewhere effect is prominent throughout (astro)particle physics and cosmology. One of the most commonly known occurrences is in collider searches for new particles, for example it was a key consideration in the Higgs boson discovery \cite{Aad:2012tfa, Chatrchyan:2012xdj}. In this example, one searches a large range of masses for a resonance, without a priori knowledge of the true mass of the particle. %, which induces the look-elsewhere effect. 
Similarly, in astrophysical searches for particles one seeks resonances in the energy flux of various astrophysical spectra, where the true energy signature of particle is unknown. Examples include: constraining the dark matter self-annihilation cross-section via
gamma ray emission from galaxy clusters \cite{Anderson_2016},
searching for
WIMPs via charged cosmic rays \cite{Reinert_2018}, searching for non-baryonic dark matter via X-ray emission from the Milky Way \cite{10.1093/pasj/psv081}, and explaining the source of high energy astrophysical neutrinos \cite{Aartsen_2014, Emig_2015}.
In terms of cosmology, the look-elsewhere effect occurs in searches for gravitational wave signals from black hole or neutron star mergers 
%in experiments such as LIGO in VIRGO 
\cite{cannon2015likelihoodratio, Abbott_2016, Messick_2017}. Here one searches large time series for a signal, where the time and shape of the event are unknown. A further cosmological example is searching for signatures of inflation in the primordial power spectrum \cite{Fergusson:2014hya, Fergusson:2014tza, Hunt_2015}. 
%The look-elsewhere effect also occurs in numerous other areas of physics and beyond, motivating the need for a fast method to account for it.

The look-elsewhere effect is also prevalent in other areas of physics and beyond, for example: in astronomy it occurs when detecting exoplanets via stellar photometry, where the period and phase of the planets' transits are unknown (e.g.~\cite{robnik2019kepler}); in biology 
it occurs when considering large DNA sequences to study
%modern DNA sampling techniques can be used to perform
genetic association 
%to find links between genotypes and phenotypes, 
%suffer from the look-elsewhere effect 
\cite{Genes1, Genes2};
%. Another medical example is the process of 
in medicine it occurs when
testing the effectiveness of drugs in clinical trials \cite{drugs}; and in theology it occurs when attempting to find hidden prophecies in religious texts \cite{BibleCode2_mckay1999}. Therefore, given the apparent ubiquity of the look-elsewhere effect, there is much motivation for a fast method to account for it.

Many simple general methods exist to mitigate for the look-elsewhere effect in the case of discrete problems, for example if one is testing multiple drugs for their effectiveness at treating a disease \cite{drugs}. The number of drugs tested, more generally known as the trials factor, quantifies the extent of the look-elsewhere effect. The larger the trials factor, i.e.~the more drugs tested, the larger the chance of a false positive arising due to a statistical fluctuation. Methods such as the Bonferroni correction \cite{bonferroni} and \v{S}id\'{a}k correction \cite{sidak} use the trials factor to correct the conclusions of a hypothesis test in light of this effect. 
%For example the Bonferonni correction increases the $p$-value by multiplying with the trials factor.
There is however no unique definition of the 
trials factor when searching a continuous parameter space for a signal.
Thus, a common, brute-force approach to account for the look-elsewhere effect for continuous problems is to perform many simulations of an experiment assuming there is no signal. One can then estimate the $p$-value of a chosen test statistic, usually related to the maximum likelihood, 
and in turn define a relation between the significance of a signal and the test statistic. This means that to conclude a detection at the 5-sigma level, corresponding to a $p$-value of order $10^{-7}$, one would need to simulate more than $\sim 10^7$ realizations of the experimental data, which is computationally expensive.
A faster method, developed in the context of high energy physics, is to approximate the asymptotic form of the $p$-value by counting upcrossings, requiring fewer simulations \cite{Gross}.
In both of these cases new simulations are required each time a new model is considered, and the simulations may not be an accurate representation of the data. 
In this paper we seek a general approach that can be directly applied to experimental data, without the need for simulations.

Our approach applies Bayesian logic to tackle the look-elsewhere effect. 
The Bayesian evidence is equal to the prior-weighted average of the likelihood over the 
parameter space, which can be considerably lower than the maximum likelihood if 
the prior is broad. This integration over the prior accounts for the look-elsewhere effect by penalizing large prior volumes. When considering large prior volumes, the likelihood is typically multimodal, with most of the peaks corresponding to noise fluctuations rather than physical sources. 
%This makes it difficult to compute the Bayesian evidence and the Bayes Factor. 
%We thus propose a fast method to compute the evidence by separately considering each peak in the posterior and using the Laplace approximation. 
In order to estimate the location of a physical signal, and its associated statistical significance, one typically considers a point estimator, such as the 
%maximum likelihood estimator (MLE), or 
maximum a posteriori (MAP) estimator which maximizes the posterior density. 
%or maximum posterior mass (MPM) which selects one peak as the most `important'.
By applying the Laplace approximation, we introduce a Bayesian generalization of the MAP estimator, referred to as the maximum posterior mass (MPM) estimator, which corrects the MAP estimator by the prior-to-posterior volume ratio. %posterior density multiplied by the posterior volume.
Then, by drawing an analogy between Bayesian and frequentist methodology, we present a hybrid of the MAP and MPM estimators, called the maximum posterior significance (MPS) estimator, which determines the most significant peak in light of the look-elsewhere effect.
%While Bayesian hypothesis testing typically has a different interpretation to frequentist $p$-values, 
The frequentist properties of the MPS estimator are shown to be independent of the look-elsewhere effect, providing a universal way to quantify the $p$-value, or statistical significance,
without the need for expensive simulations.

The outline of this paper is as follows. In section \ref{sec:MPM} we review Bayesian posterior inference and hypothesis testing for a multimodal posterior, by discussing MAP estimation and then introducing MPM estimation.
We then draw an analogy between Bayesian and frequentist philosophy in section \ref{sec:bayes_freq} to motivate MPS estimation as the appropriate technique to tackle the look-elsewhere effect. The following three sections then apply this method to various examples:
section \ref{sec:HEP} considers a resonance search, which can be thought of as a toy example of a collider or astrophysical particle search; 
section \ref{sec:white} considers a white noise time series, which can be thought of as a toy example of a gravitational wave search;
and section \ref{sec:planck} considers a search for non-Gaussian models of cosmological inflation using Planck data \cite{Planck13_L}. Note that 
%the examples of 
section \ref{sec:HEP} is the main example, as it illustrates the key advantages of MPS,
%(which apply to all examples), 
with the other examples complementary. 
%Section \ref{sec:white} complements section \ref{sec:HEP} by exploring the effects of model complexity. 
Finally, we summarize and conclude in section \ref{sec:conclusion}.

\section{Bayesian posterior inference and hypothesis testing}
\label{sec:MPM}

Two of the main tasks of Bayesian statistical 
analysis are posterior inference and hypothesis 
testing. Consider 
 a model with parameters $\bi{z}= \{z_j\}_{j=1}^M$, and 
%we seek the posterior of $\bi{z}$ given 
data $\bi{x}= \{x_i\}_{i=1}^{N_d}$ that depends on $\bi{z}$.
The inference of $\bi{z}$
%, given $\bi{x}$, 
is given by its posterior
\begin{equation}
    p(\bi{z}|\bi{x})
    =\frac{p(\bi{x},\bi{z})}{p(\bi{x})}
    =\frac{p(\bi{x}|\bi{z})p(\bi{z})}{p(\bi{x})},
    \label{loss}
\end{equation}
where $p(\bi{x}|\bi{z})$ is the likelihood of the data, $p(\bi{z})$ is the prior of $\bi{z}$,
and $p(\bi{x})=\int d\bi{z} ~p(\bi{x}|\bi{z})p(\bi{z})$ is the
Bayesian evidence, also known as the normalization, marginal likelihood, or
partition function. Typically, one can evaluate the joint probability $p(\bi{x},\bi{z})$, but not the evidence, which 
makes the posterior inference analytically intractable. 
This is usually handled using simple approximations or 
 Monte Carlo Markov Chain methods \cite{Chen2000}. 

A related problem is that of a hypothesis testing. 
In this case there are two different hypotheses,
$H$ and $H_0$, each with their own model parameters,  $\bi{z}$ and $\bi{z}_0$.  
In Bayesian methodology, hypothesis testing is performed
using the Bayesian evidence ratio of the two hypotheses,
which 
gives the Bayes factor
\begin{equation}
    B 
    \equiv \frac{p(\bi{x}|H)}{p(\bi{x}|H_0)}, 
    \label{BF}
\end{equation}
where the Bayesian evidence for hypothesis $H$ is given by
\begin{equation}
    p(\bi{x}| H) 
    = \int d\bi{z} ~p(\bi{x}|\bi{z},H) p(\bi{z}|H).
    \label{ev}
\end{equation}
%%%The dimensionality of $\bi{z}$ can differ between the two hypotheses. In this paper we will be concerned with problems in which the the parameters of $H_0$ are a subset of the parameters of $H$.
%
The Bayesian evidence and Bayes factor are also 
analytically intractable and  
harder 
to evaluate than posteriors, especially for high dimensional $\bi{z}$, 
although recent numerical methods such as 
Gaussianized Bridge Sampling \cite{HeSeljak} have 
made the problem easier. For the sake of 
exposition we will not consider such 
methods in this work, but instead use 
analytical approximations that give the Bayes factor 
an intuitive meaning. It is worth keeping in 
mind however that the full Bayes factor calculation 
can always be performed numerically, without 
any approximations. 

\subsection{Maximum a Posteriori (MAP) estimation}
\label{sec:MAP}

Given the analytical intractability of 
posterior inference and hypothesis testing, 
one often chooses an estimator to extract useful information from the posterior.
A common estimator is the maximum a posteriori (MAP) point estimator, which corresponds to the global maximum of the posterior.
If the prior is flat, as it will always be in this paper, this equals the maximum likelihood estimator (MLE), 
which maximizes the likelihood. 
Mathematically, MAP is defined via 
\begin{equation}
 {\rm MAP}:   \argmax_{\bi{ z}}p(\bi{z} | \bi{x}).
 \label{MAP}
\end{equation}
%
%
%If the solution is inside the interval of $\bi{z}$ and $\bi{s}$ then one can search for it using MAP equation $\nabla_{\bi{z}} \mathcal{L}_p=\nabla_{\bi{s}} \mathcal{L}_p=0$.
%
For the purpose of comparing data to a null hypothesis, a useful quantity to define is
\begin{equation}
    q_L(\bi{z}) \equiv 2 \ln \frac{p(\bi{x}|\bi{z})}{p(\bi{x}|\bi{z}_0)},
    \label{qL}
\end{equation}    
where $\bi{z}_0$ represents the values of the parameters under the null hypothesis, and a subscript of $L$ is used because the argument of the logarithm is the Likelihood ratio. 
To assess the significance of a result one considers the maximum value of $q_L$, 
%denoted $\hat{q}_L$, 
which in the case of a flat prior is equal to $q_L$ evaluated at the MAP: $\hat{q}_L = q_L(\bi{z}_{\rm MAP})$.
For a Gaussian likelihood, this is equal to the chi-squared ($\chi^2$), and in the absence of the look-elsewhere effect $\sqrt{\hat{q}_L}$ typically gives the statistical significance. However, we will see that this test statistic greatly suffers from the look-elsewhere effect.% due to the a posteriori nature of evaluating at the MAP.

\subsection{Maximum Posterior Mass (MPM) estimation}

MAP is often a
good point estimator in low dimensions if there is a single mode in the posterior. However, if the posterior has
several modes, a more reasonable 
point estimator associates with the highest posterior mass.
%, defined as the integral of the posterior over the region of the mode. 
We refer to this as
the maximum posterior mass (MPM) estimator.

For the purposes of this work, we will consider the example of a multimodal posterior consisting of a 
sum of multivariate Gaussian distributions; this has been shown to be a good approximation in many practical cases \cite{SeljakYu2019}.
%\footnote{Sometimes the posterior peaks are not well described by  multivariate gaussians. This happens particularly often  when the peak is close to the boundary set by the prior.  Methods have been developed to address such situations  \cite{SeljakYu2019}.}. 
We thus consider a posterior of the following form,
\begin{equation}
    p(\bi{z}|\bi{x})
    %\approx q(\bi{z})
    %\equiv \exp(-\mathcal{L}_q)
    = \sum_l w^l N({\bi{z}};\bi{\mu}^l, \bi{\Sigma}^l),
    %\equiv \sum_l w^l N^l(\bi{z}),
\label{gm}
\end{equation}
where $N(\bi{z};{\bi{\mu}}, {\bi{\Sigma}})$ is a multivariate normal distribution with mean $\bi{\mu}$ and covariance matrix $\bi{\Sigma}$, and the data dependence has been dropped for neatness. 
Note that working with a posterior of this form is equivalent to applying the Laplace approximation to a general multimodal posterior in the upcoming derivations.
In this model, the mass of mode $l$ is proportional to the weight $w^l$, which is normalized such that $\sum_l w^l = 1$.
%and the MPM point estimator corresponds to the mode with the largest weight.

Assuming that only one component contributes at each peak, the weight of mode $l$ is given by
evaluating the posterior at the location of the mode, $\bi{z}=\bi{\mu}^l$,
%such that the second term is zero and we are only left with the volume term $\ln \det \bi{\Sigma}^l$, giving
\begin{equation}
    \ln w^l 
    = \ln p(\bi{\mu}^l|\bi{x}) - \ln N({\bi{\mu}^l};\bi{\mu}^l, \bi{\Sigma}^l)
    = \ln p(\bi{\mu}^l|\bi{x}) +\frac{1}{2}\left[\ln \det \bi{\Sigma}^l+M\ln(2\pi)\right].
    \label{wl}
\end{equation}
To obtain a quantity that can be readily computed, we multiply each weight by the normalization $p(\bi{x})$ to give the mass $m^l$, defined by
\begin{equation}
    \ln m^l 
    \equiv \ln w^l + \ln p(\bi{x}) 
    = \ln p(\bi{x}|\bi{\mu}^l)+\ln p(\bi{\mu}^l)+\frac{1}{2}\left[\ln  \det \bi{\Sigma}^l +M\ln(2\pi)\right].
    \label{ml}
\end{equation}
%
%The log mass of each mode is %, up to a constant, 
%equal to the log likelihood plus the log prior density plus the log of the posterior volume at the peak, which is approximated with the log determinant of the covariance matrix $  \bi{\Sigma}^l$.
Thus, the mass of each mode is %, up to a constant, 
equal to the log likelihood multiplied by the product of the prior density and the posterior volume at the peak, where the posterior volume is defined as 
\begin{equation}
V_{\rm posterior} \equiv (2 \pi)^{M/2} \sqrt{\det \bi{\Sigma}}.
\label{Vpost}
\end{equation}
%For a monomodal posterior, the product of the prior density and the posterior volume is often referred to as the Occam's razor penalty, as it penalizes the likelihood for model complexity \cite{MacKay2003}.

\begin{figure}[t]
\centering \includegraphics[height=0.32\textwidth]{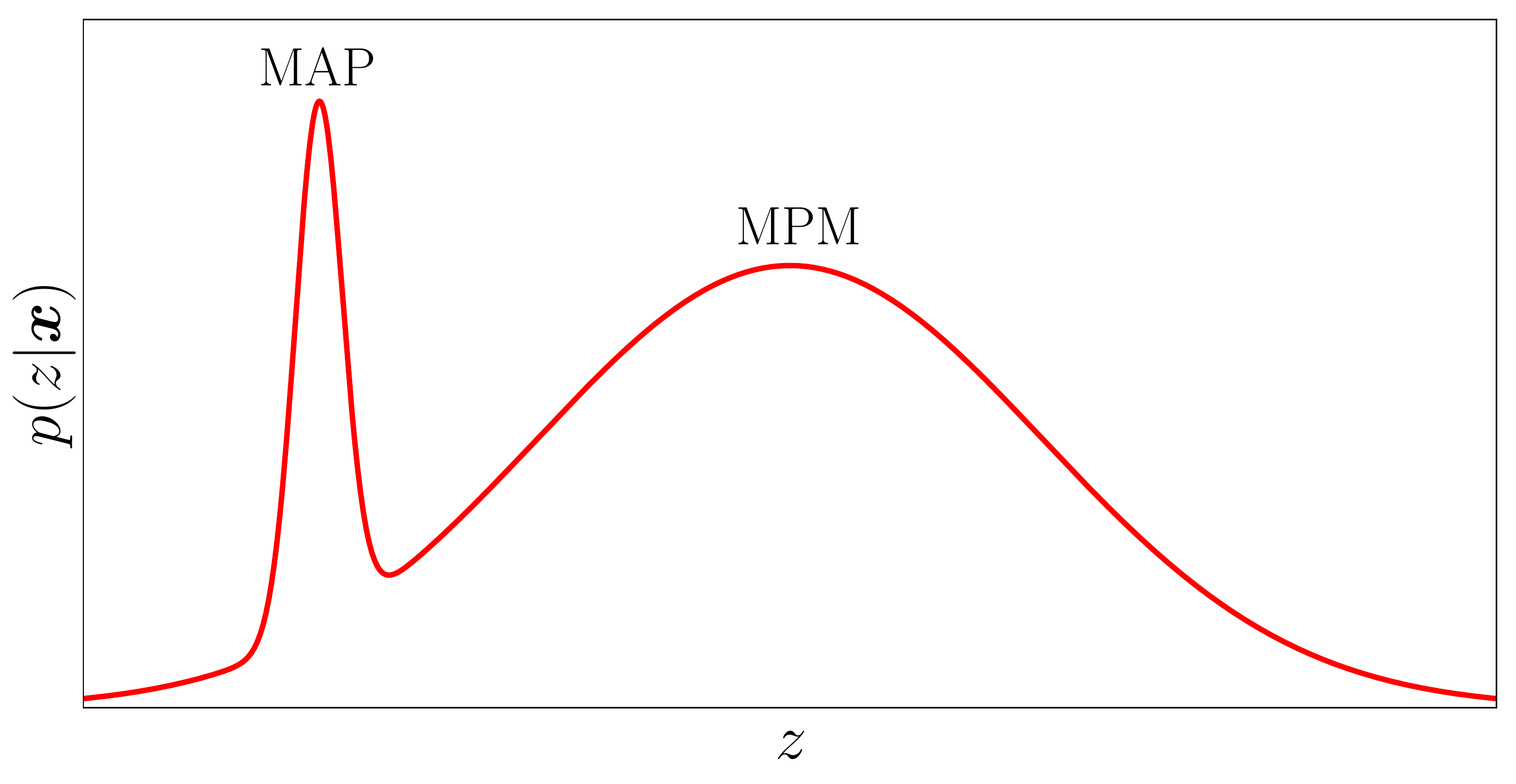} 
\caption{Plot of a bimodal Gaussian posterior for a 1d example in which 90\% of the posterior mass is assigned to the right peak and 10\% to the left. MPM yields the mode that maximizes the posterior mass and is close to the true mean, whereas MAP maximizes the posterior density and can be distant from the mean and represent only a small 
fraction of posterior mass. 
%For the Bayes factor of section \ref{sec25} the plot shows the 2-d posterior of $z_1$, evaluated at the peak value $\mu_{z_2}$.
%FIX: $z_1$ confusing.
}
\label{fig:multimodal}
\end{figure}

The MPM estimator corresponds to the mode with the  highest mass, hence to determine the MPM mode one would compute the $\ln m^l$ by first finding the positions
of all local posterior maxima $\bi{\mu}^l$, and then computing $\bi{\Sigma}^{l}$ using the inverse of the Hessian at each peak.  
Qualitatively, MPM corresponds to maximizing the posterior density multiplied by the posterior volume $\sim \sqrt{\det \bi{\Sigma}}$, whereas MAP only maximizes the former.
It is apparent that if there are multiple modes in the posterior, the one 
that has the largest posterior mass does not necessarily have the 
largest posterior density, as shown in figure 
\ref{fig:multimodal}.
%If there are multiple modes we need to find 
%the one that corresponds to the global minimum of MPM. 
In some situations the 
MPM mode will dominate the posterior mass
such that
%, even  when MPM does not correspond to MAP. 
 %
%While it is always best to report
%the full posterior, 
%or its moments, e.g. the mean is 
%\begin{equation}
% $   \E[\bi{\mu}]=\sum_l w^l \bi{\mu}^l$,
%\end{equation}
%(similar expressions can be obtained for the higher moments), 
%In such a case
the MPM mode alone gives a useful way to summarize the posterior.
%in the limit of a single peak dominating the posterior mass. 

 %, or equivalently, maximizes the posterior mass given by the weight $w^l$.
%if the sampling function $\tilde{p}$ is a sum of delta functions.  
%these solutions correspond to multimodal MAP solutions 
%where the gradient vanishes, and the Hessian $\bi{\Sigma}^{-1}$
%is given by the Laplace approximation at the mode. 

\subsection{Hypothesis testing with MPM}

Consider a model with parameters 
$z_1,z_2, ..., z_M$, with $z_1$ corresponding to the amplitude of a feature, and $\bi{z}_{>1}$ corresponding to the properties of the feature. For example, $z_1$ might correspond to the amplitude of a signal detected in a time series at time $z_2$.
A typical analysis would scan over the $\bi{z}_{>1}$, finding the 
best fit value for the amplitude $z_1$ at each point, giving rise to a multimodal posterior.

In this work we wish to determine whether or not a dataset contains a true anomaly. In the language of hypothesis testing, we wish to compare the hypothesis that there is an anomaly $H$, corresponding to $z_1 > 0$, to the null hypothesis that there is no anomaly $H_0$, corresponding to $z_1=0$. We assume the common case that the parameters of $H_0$ are a subset of the parameters of $H$, with $H$ reducing to $H_0$ when $z_1 = 0$. There may also be parameters other than $\bi{z}$ that are common to both models, but these are of secondary importance when considering the look-elsewhere effect and we drop these from the notation.

Using equation \ref{ml} with $\sum w^l=1$ implies that the Bayesian evidence for hypothesis $H$ is given by
\begin{equation}
    p(\bi{x}| H) 
    = \int d\bi{z} ~p(\bi{x}|\bi{z},H) p(\bi{z}|H)
    = \sum_l m^l,
    \label{ev_sum_m}
\end{equation}
where the $m^l$ correspond to the masses under  hypothesis $H$.
Hence, each mode contributes its mass to the evidence. It follows that the mass of mode $l$ corresponds to the Laplace approximation of the evidence integral in equation \ref{ev_sum_m}, integrated over the region of the mode.
Because the null hypothesis does not depend on $\bi{z}_{>1}$,
%\footnote{There may be parameters shared by both hypotheses, but assuming they are independent of $\bi{z}$, their effect cancels out in Bayes factor.}
the evidence for the null hypothesis is given by the likelihood evaluated at $z_1=0$, that is  $p(x | H_0) = p(x | z_1 = 0) \equiv p_0(x)$.
Together with equation \ref{ev_sum_m} this gives the Bayes factor
\begin{equation}
    B
    \equiv \frac{p(\bi{x}|H)}{p(\bi{x}|H_0)}
    = \frac{1}{ p_0(\bi{x})}  \sum_l m^l
    \equiv \sum_l b^l,
    \label{BFgen}
\end{equation}
where $b^l$ is defined as the contribution of mode $l$ to the Bayes factor. Using equation \ref{ml} gives
\begin{equation}
    b^l 
    = \frac{p(\bi{x}|\bi{\mu}^l)}{ p_0(\bi{x}) } p(\bi{\mu}^l) (2 \pi)^{M/2} \sqrt{\det \bi{\Sigma}^l}
    = \frac{p(\bi{x}|\bi{\mu}^l)}{ p_0(\bi{x}) } \frac{ V_{\rm posterior} (\bi{\mu}^l) }{ V_{\rm prior} (\bi{\mu}^l) },
    \label{bl}
\end{equation}
%, and we have dropped $s$ subscripts to avoid clutter. 
where we have introduced the effective volume of the prior at $\bi{\mu}^l$ as,
\begin{equation}
    V_{\rm prior}^{-1}(\bi{\mu}^l) \equiv p(\bi{\mu}^l),
    \label{vp}
\end{equation}
appropriate for the case of a narrow posterior relative to the prior. 
In the remainder of this paper we will drop the $\bi{\mu}^l$ dependence of 
the prior volume, as appropriate for a flat prior on $\bi{z}$.

Intuitively, one can think of each $b^l$ as the Bayes factor one would get if mode $l$ were the only mode in the posterior. If the maximum $b^l$ is sufficiently large, it alone can provide a useful approximation to the Bayes factor, meaning the MPM mode dominates the Bayes factor.
The first ratio on the right hand side of equation \ref{bl} corresponds to the likelihood ratio of the signal hypothesis to the
null hypothesis, evaluated at the location of the peak, $\bi{z}=\bi{\mu}^l$. This is greater than or equal to $1$ since adding 
 parameters to the null hypothesis 
can only improve the fit. The second ratio gives the ratio of the posterior volume
to the prior volume at the peak, which is always less than $1$. 
%because posterior volume is always smaller than prior volume. 
This acts as a penalty to the likelihood ratio, often referred to as the Occam's razor penalty \cite{MacKay2003}, 
or model complexity penalty, and compensates for 
the look-elsewhere effect in the case of a multimodal posterior. 
The higher the prior-to-posterior volume ratio, the higher the chance that peaks with a high likelihood will occur because of statistical fluctuations, thus the larger the penalty required to compensate.

Just as $q_L$ is the estimator associated with MAP, we can define  $q_b \equiv 2 \ln b$ as the estimator associated with MPM, such that
\begin{equation}
    q_b
    =  q_L  - 2\ln \frac{V_{\rm prior}}{V_{\rm posterior}}.
    \label{qbhat}
\end{equation}
The MPM mode corresponds to the mode with maximum $q_b$.
This illustrates how the MAP estimator ignores the look-elsewhere penalty by effectively considering the posterior and prior to be overlapping delta functions, which presumes a priori knowledge of the parameters and gives a prior-to-posterior volume ratio of unity. %Note that the hat on $V_{\rm posterior}$ signifies that it is the posterior volume of the MPM mode; we will now drop this notation for neatness.

An interesting question to consider is whether one can relate $q_b$ to the look-elsewhere corrected statistical significance in a frequentist sense. In the absence of the look-elsewhere effect, the significance is given by $\sqrt{q_L}$, but simply taking $\sqrt{q_b}$ as the look-elsewhere corrected significance would not be correct. In the next section we turn to a frequentist description of the look-elsewhere effect to motivate a new estimator which applies a small modification to $q_b$ and has a simple interpretation in terms of the significance, or $p$-value.

 %\textcolor{red}{We use hats to denote quantities associated with the MPM -- this will apply throughout the paper, except when $\hat{q}_L$ appears alone, in which case it refers to the maximum $q_L$, i.e. $q_L$ at MAP, as in section \ref{sec:MAP}. While this notation is somewhat sloppy, it keeps things neat; it should be apparent what a hat signifies based on the context.}

Before ending this section we discuss the choice of priors appropriate for a look-elsewhere analysis.
If one has no prior knowledge regarding the location of an anomaly, then a uniform prior for the $\bi{z}_{>1}$
parameters is appropriate. If 
the prior is wide and posterior narrow this 
induces a large look-elsewhere effect. This choice of prior is not controversial. On the other hand, the choice of prior for the
amplitude parameter $z_1$ is less clear. If one has no prior knowledge of the signal amplitude, then one should be open to a signal of any size, however one does not want the amplitude prior to
induce a look-elsewhere penalty. 
In Bayesian hypothesis testing the amplitude parameter is treated analogously to the other parameters, thus if one uses too broad an amplitude prior it will induce an unwanted look-elsewhere penalty, whereas if one chooses too narrow an amplitude prior one risks discounting a large signal. 
Based on this we rewrite $b$ in the following form, explicitly separating the marginalization over $\bi{z}_{>1}$ and $z_1$, 
\begin{equation}
    b =  e^{q_L/2} \frac{V_{\rm >1, posterior}}{V_{\rm >1, prior}} \frac{V_{\rm 1,  posterior}}{V_{\rm 1, prior}}.
    \label{bhat_sept}
\end{equation}
The posterior volume terms are given by the covariance matrix, as in equation \ref{Vpost}, and $V_{\rm >1, prior}$ is given by the choice of prior on $\bi{z}_{>1}$.
It thus remains to justify a choice of $V_{\rm 1, prior}$, which we will do by turning to a frequentist description of the look-elsewhere effect in the next section.

%%In summary, we have shown that approximate marginalization around the maximum posterior mass peak gives a fast way to evaluate the Bayes factor. Furthermore, we have shown that, unlike the likelihood ratio, the Bayes factor incorporates a look-elsewhere penalty related to the prior-to-posterior volume ratio.% In the next section we will apply this to define a new estimator to quantify the significance of a hypothesis test in light of the look-elsewhere effect.

\section{From Bayesian to frequentist hypothesis 
testing
%: Maximum Posterior Significance (MPS)
}
\label{sec:bayes_freq}

Standard statistics literature states that Bayesian and frequentist 
hypothesis testing follow different methodologies 
and may give very different results. One famous illustration of this is the Jeffreys-Lindley ``paradox'' \cite{Lindley}, however, there is much debate as to whether this is indeed a paradox and how relevant it is for scientific discourse (see \cite{Cousins_2014} for a review in the context of high energy physics).
While 
Bayesian statistics uses the Bayes factor for hypothesis 
testing, 
frequentist statistics uses the maximum likelihood ratio, or $\hat{q}_L$. %, for problems with a single degree of freedom.
One of the most important aspects of frequentist 
methodology is the computation of the false positive
rate using the $p$-value, which quantifies how often a test statistic, for example $\hat{q}_L$, will take a specific value  
or larger under the assumptions of the null hypothesis. This 
has an intuitive interpretation as it directly relates 
to the false positive rate of the test statistic.
%, while  intrinsically incorporating model dependence. 
On the other hand, Bayesian methodology rejects
the $p$-value.
The basis for this rejection is 
the likelihood principle, which states that any inference 
about the parameters $\bi{z}$ from the data $\bi{x}$ 
can only be made via the likelihood $p(\bi{x}|\bi{z})$. When the likelihood principle is 
applied to testing a hypothesis with parameters $\bi{z}$ 
one must 
use the marginal likelihood by integrating out 
these parameters -- as in the Bayesian evidence of equation \ref{ev} -- thus Bayesian methodology explicitly satisfies 
the likelihood principle. 
It is commonly argued that $p$-values 
violate the likelihood principle, because they 
rely on the frequentist properties of a distribution 
that go beyond the likelihood principle. 
However, the Bayes factor provides a less reliable tool for model comparison, as it is often interpreted in terms of arbitrary, model-independent scales
%, such as the Jeffreys' scale 
\cite{Nesseris_2013}, unlike the $p$-value which directly relates to the false positive rate.
%Here we 
%will show that the two can be heuristically related with a specific choice of prior. 

We seek to elucidate how the 
answers of the two schools of statistics 
relate to one and other when it comes to the hypothesis 
testing.
Both schools of statistics should give a 
similar, or at least related
answer, when the question is phrased
similarly. For uncertainty quantification it is 
often argued that the two schools do not answer 
the same question, since the Bayesian school treats data
as fixed and varies the models, while the frequentist school 
varies the data at a fixed model. However, when it 
comes to hypothesis testing the distinction is less 
prominent: for example, when comparing two 
discrete hypotheses without any marginalizations, the 
answer in both cases gives the likelihood ratio as the optimal statistic (assuming equal prior for the two hypotheses). For 
continuous hypotheses it is often argued this is not 
possible. 
Here we will show that the two answers, 
the $p$-value and the Bayes factor, can be  related with a specific choice of prior. 
It is important to emphasize that we are not claiming to equate the Bayesian and frequentist methodologies, but rather  motivate a connection. 
%In frequentist 
%methodology the $p$-value must be compared to some 
%predetermined confidence level $\alpha$, rejecting the 
%null hypothesis if below it. 

%In this work we use the typical definition of the $p$-value: 
In this work we define the $p$-value as the probability under the null hypothesis, $H_0$, of a random variable, $Q$, to be observed to have a value equal to or more extreme than the value observed, $q$. We thus use the notation $P(Q \geq q)$ for the $p$-value. 
%We emphasize this to distinguish from another quantity that is often referred to as the $p$-value: the probability of observing a value $q$ that is larger than or equal to some fixed threshold $q'$.
%
To compute the $p$-value of a test statistic, one must consider how the test statistic is distributed under the null hypothesis. For the example of $\hat{q}_L$ this distribution is not universal: scanning over continuous variables, as in the look-elsewhere effect, will modify this distribution in a model dependent manner. Moreover, increasing the model complexity in other ways, for example by including extra degrees of freedom, will further modify the distribution. To account for extra degrees of freedom, Wilks' theorem \cite{wilks1938} 
provides the asymptotic distribution of $\hat{q}_L$ for a hypothesis test where $H$ has $\nu$ more degrees of freedom than $H_0$.
%, as a chi-squared with $\nu$ degrees of freedom. 
However, Wilks' theorem
relies on technical conditions, such as the observed value not being at the 
edge of the interval, and 
does not consider the look-elsewhere
effect. Generalization of Wilks' theorem for the look-elsewhere effect have been considered in \cite{Davies, Davies2} and have been translated into a practical procedure in \cite{Gross}. As a result, a frequentist approach consists of a series of
considerations to determine the change in the distribution of $\hat{q}_L$ due to different sources of model complexity. This is unlike the
Bayesian methodology where all forms of model complexity are accounted for in the same way, as they are encoded into the Bayes factor.
By connecting the two methodologies, we will present a test statistic whose distribution is universal, regardless of the model complexity and look-elsewhere effect. %Thus, considering the $p$-value of this test statistic provides a universal prescription to account for the look-elsewhere effect without the need for simulations. 

%Since Bayes factor can always be computed and handles 
%all the effects we will advocate the use of Bayes factor 
%for the look-elsewhere
%analysis, converting it to a $p$-value to quantify 
%the false positive rate. 

\subsection{Maximum Posterior Significance (MPS) estimation}
\label{sec:MPS}

We start by considering the typical case of one degree of freedom, corresponding to a single signal with amplitude $z_1$ and features described by $\bi{z}_{>1}$.
We denote $q_L$ maximized over the amplitude parameter only as $\check{q}_L (\bi{z}_{>1}) \equiv \max_{z_1} q_L(\bi{z})$, not to be confused with $\hat{q}_L \equiv \max_{\bi{z}} q_L(\bi{z})$ which is $q_L$ maximized over all parameters. %We will use uppercase $Q$ to denote random variables.
For a $t$-tailed test (where $t$ is equal 1 or 2), Wilks' theorem gives the asymptotic $p$-value of $\check{q}_L$, at any position $\bi{z}_{>1}$, as
\begin{equation}
    P(\check{Q}_L \geq \check{q}_L) = \frac{t}{2} \tilde{F}_1(\check{q}_L) \xrightarrow[]{\check{q}_L \rightarrow \infty}  \frac{t}{\sqrt{2 \pi \check{q}_L}}  e^{-\check{q}_L/2},
    \label{pql_local}
\end{equation}
where $\tilde{F}_\nu$ is the complementary cumulative distribution function (CCDF) of a chi-squared random variable with $\nu$ degrees of freedom. This maximization over $z_1$ at a fixed choice of $\bi{z}_{>1}$ corresponds to the $p$-value in the absence of the look-elsewhere effect, referred to as the \textit{local} $p$-value.
Further maximizing over $\bi{z}_{>1}$ introduces the look-elsewhere effect, which can be parameterized by multiplying by the trials factor $N$ such that
\begin{equation}
    P(\hat{Q}_L \geq \hat{q}_L) = N \frac{t}{\sqrt{2 \pi \hat{q}_L}}  e^{-\hat{q}_L/2}.
    \label{pql_global}
\end{equation}
This is referred to as the \textit{global} $p$-value.
It is this form that leads to the Bonferroni correction \cite{bonferroni} which divides the type I error by $N$ to account for the look-elsewhere effect.
For discrete problems the trials factor equals the number of trials performed. However, in the continuous case it is ill-defined, but it quantifies how the probability of finding a spurious peak increases as one looks elsewhere in the space spanned by $\bi{z}_{>1}$. Accounting for the look-elsewhere effect thus requires an expression for the trials factor.
%we propose is given by the prior-to-posterior volume for $\bi{z}_{>1}$ evaluated at the most significant mode.
%Intuitively one can think of the number of trials as the number posterior volumes that fit within the prior volume.
%We will now motivate this proposal more concretely.

%
It follows from equation \ref{pql_global} that one can define a test statistic,
\begin{align}
    q_S = q_L - 2 \ln N + \ln 2 \pi q_L - 2 \ln t
    \label{qa}
\end{align}
such that the global $p$-value tends to
\begin{equation}
    P(\hat{Q}_S \geq \hat{q}_S) \rightarrow e^{-\hat{q}_S/2},
    \label{pqa}
\end{equation}
as either $N \rightarrow \infty$ or $\hat{q}_S \rightarrow \infty$, so this also applies for $N=1$. 
See Appendix \ref{app:qa} for a derivation. 
Unlike $\hat{q}_L$, $\hat{q}_S$ has a distribution that is independent of $N$ -- the look-elsewhere effect has been absorbed into the test statistic.
Intuitively one can think of the $2 \ln N$ term as a penalty to $q_L$ to correct for the look-elsewhere effect, while the  $\ln 2 \pi q_L$ term removes $q_L$ dependent bias, ensuring the $p$-value depends on $\hat{q}_S$ alone in the asymptotic limit.  
Thus to account for the look-elsewhere effect one need only compute $\hat{q}_S$ and use this equation to compute the $p$-value. 
%Intuitively one can think of each peak as having a quantity $e^{-q_S/2}$ associated with it. 
Because the $p$-value is a monotonically decreasing function of $\hat{q}_S$, one can think of selecting the peak with maximum $q_S$ as selecting the peak with minimum $p$-value or maximum statistical significance. % (in a frequentist sense).
We refer to the mode with maximum $q_S$ as the MPS mode, deferring an  explanation for this nomenclature until the end of the subsection.
The similarity of $q_S$ to $q_b$ from equation \ref{qbhat} suggests a connection between the frequentist and Bayesian pictures, and we now invoke this connection to find an expression for $N$ and in turn generalize the Bonferroni correction to continuous parameters.
%However, in practise this method is not practised by frequentists, the reason is that the trials factor is ill defined for continuous parameters. 
%However,
%%The Bayesian methodology gives a natural interpretation for the trials-factor by considering the marginal likelihood over $z_1$, enabling the use of equation \ref{pqa} to compute the $p$-value and generalizing the Bonferroni correction to continuous variables.

Heuristically, 
the Bayes factor describes the probability of the
alternative hypothesis relative to the null, 
determined by the likelihood (as measured by $\hat{q}_L$), 
while the $p$-value averages its inverse over all values 
larger than $\hat{q}_L$ and will be smaller than the 
likelihood. We expect that for higher $\hat{q}_L$ the effect is 
larger because we are further into the tail of the 
distribution. 
There is no unique relation between the two, 
but one simple option is that the $p$-value scales as $B^{-1}/\hat{q}_L \approx \hat{b}^{-1}/\hat{q}_L$, where hats now indicate quantities associated with the MPS mode. 
Because we have 
the freedom to choose the prior on $z_1$,
we can {\it define} the relation between the Bayes 
factor and $p$-value as %, which we will choose to be
\begin{equation}
%\frac{1}{2}\int_{\hat{q}_L}^{\infty} e^{-x/2}dx=\frac{e^{-\hat{q}_L/2}}{\hat{q}_L}= 
\frac{\hat{b}^{-1}}{\hat{q}_L} \equiv P(\hat{Q}_L \geq \hat{q}_L) .  
\label{pb}
\end{equation}
Comparing equation \ref{bhat_sept} with equation \ref{pql_global} then gives
\begin{align}
  %  \frac{\hat{b}^{-1}}{\hat{q}_L} &\sim P(\hat{Q}_L \geq \hat{q}_L) \\
    \frac{V_{\rm >1,prior}}{\hat{V}_{\rm >1,posterior}}
    \frac{V_{\rm 1,prior}}{\hat{V}_{\rm 1,posterior}} 
    \frac{e^{-\hat{q}_L/2}}{\hat{q}_L} &= 
     N \frac{t}{\sqrt{2 \pi \hat{q}_L}}  e^{-\hat{q}_L/2}.
    \label{Bq}
\end{align}
By requiring that this relation holds in the absence of the look-elsewhere effect, the trials factor can be identified as
\begin{equation}
N = \frac{V_{\rm >1,prior}}{\hat{V}_{\rm >1,posterior}},
\label{N}
\end{equation}
and the amplitude prior volume is given by 
%To gain intuition for what amplitude prior this corresponds to, we promote the scaling of \ref{Bq} to an equality, giving
\begin{equation}
V_{\rm 1,prior}
%=p(z_1|H)^{-1}
=  t \sqrt{\hat{q}_L} \frac{\hat{V}_{\rm 1, posterior}}{ \sqrt{2 \pi} }
=  t \sqrt{\hat{q}_L}  \hat{\sigma}_1 
\approx t\hat{\mu}_1.   
\label{z1p}
\end{equation}
In the final steps we used $\hat{V}_{\rm 1, posterior} = \sqrt{2 \pi} \hat{\sigma}_1$, where $\hat{\sigma}_1$ is the error on the amplitude parameter, $\hat{\mu}_1$, and that the signal-to-noise ratio obeys $\sqrt{\hat{q}_L} \approx \hat{\mu}_1 / \hat{\sigma}_1 $.
%This choice of prior ensures the $p$-value equals $B^{-1}/\hat{q}_L$ regardless of there being a look-elsewhere effect. 
%%%In the case of $t=2$, the prior would be symmetric about $z_1=0$, with volume $\mu_1$ on either side. 
%
Since the look-elsewhere effect leads to large $\hat{q}_L$, this 
prior volume will be larger than the posterior volume. This choice of amplitude prior volume ensures that there is no trials factor associated with the amplitude, as intuition would dictate.
Substituting equations \ref{N} and \ref{z1p} into equation \ref{qa} yields %$ \hat{q}_S 
%= \hat{q}_L - 2 \ln \frac{V_{\rm >1,prior}}{V_{\rm >1,posterior}} + \ln 2 \pi \hat{q}_L - 2 \ln t 
%= \hat{q}_b + 2 \ln \hat{q}_L$, 
\begin{align}
    q_S 
    %=  q_L - 2 \ln \frac{V_{\rm >1,prior}}{V_{\rm >1,posterior}} + \ln 2 \pi q_L - 2 \ln t 
    = q_b + 2 \ln q_L.
    \label{qa2}
\end{align}
Hence, we have effectively applied a modification to the MPM estimator to give a combination of the MPM and MAP estimators, so that the asymptotic $p$-value is neatly given by $e^{-\hat{q}_S / 2}$. In the context of the look-elsewhere effect, the mode with maximum $q_b$ will typically also be the mode with maximum $q_L$, and thus maximum $q_S$. However, this equivalence of MAP and MPM may not always be the case, as shown in figure \ref{fig:multimodal}. 

A pure Bayesian might argue that equation \ref{z1p} is not a valid 
prior, since it depends on the a posteriori amplitude parameter
%of the MPS mode
$\hat{\mu}_1$; however, this prior does have an intuitive justification. 
%one can also argue that 
If a scientist is willing to consider a signal of any amplitude, 
%the measured value $\hat{\mu}_1$ as acceptable, 
%it means 
the 
prior cannot be zero at $\hat{\mu}_1$, as it would not make sense to discard the signal. On the 
other hand, making the prior significantly broader than $\hat{\mu}_1$ implies
the scientist has some additional information on the 
nature of the amplitude. 
%The purpose of this paper is 
%to discuss the look-elsewhere effect where the priors on 
%some parameters 
%are very broad, but they are also well justified. 
When there is no justification for broadening 
the prior, the narrowest possible prior still 
consistent with the measured value can be more 
reasonable than arbitrarily fixing the size of the prior a priori. 
%The purpose of this section is to 
%relate the Bayes factor to the $p$-value to determine $N$, not to argue what the most appropriate choice of prior for the amplitude is.
This choice of amplitude prior is simply designed to allow for a signal with any amplitude, without inducing an unwanted look-elsewhere penalty.

Note that the explicit dependence on $\hat{q}_L$ and the marginal likelihood, via $\hat{b}$, in 
equation \ref{pb} is what makes the $p$-value inconsistent 
with the likelihood principle. One could instead consider
equating $\hat{b}^{-1}$ directly with the $p$-value, making it 
consistent with the likelihood principle. This would 
require an amplitude prior of $V_{\rm 1,prior}=t\hat{\sigma}_1^2/\hat{\mu}_1$, which we deem unreasonable as it is smaller than the posterior volume $\hat{V}_{\rm 1, posterior}$. 
We emphasize that the equality of $\hat{b}^{-1}/\hat{q}_L$ to the $p$-value is not strictly required for our approach to the look-elsewhere effect, but provides intuition for the Bayesian-frequentist connection. At its core, our method considers the test statistic $\hat{q}_S$, from equation \ref{qa}, and replaces the trials factor $N$ with the prior-to-posterior volume of the non-amplitude parameters $\bi{z}_{>1}$.
Intuitively one can think of the number of trials as the number posterior volumes that fit within the prior volume.
%; the rest of the above discussion is ancillary.
%%%A Bayesian disturbed by this prior could discard this discussion and instead view our method as using the test-statistic $\hat{q}_S$ with $N$ as the prior-to-posterior volume of the non-amplitude parameters. The equality of $B^{-1}/\hat{q}_L$ to the $p$-value is not needed to tackle the look-elsewhere effect, but we deem it insightful to draw the analogy.
%
%It is often argued that broad priors are conservative, 
%but it is just the opposite: broad priors need a
%theoretical justification, and in the absence of it 
%the narrowest possible prior consistent with the analysis
%should be adopted. 
%

Because the asymptotic $p$-value scales linearly with the prior volume, the non-asymptotic form of the $p$-value can be derived by
dividing the prior volume into $K\gg 1$ 
regions and evaluating the $p$-value for each. 
Assuming independence between these regions, the product of the $p$-values for each region can be used to 
obtain $p$-value of the full volume. Further assuming that the asymptotic regime still applies, this gives %Specifically, 
%if we want to evaluate $p$-value at a
%prior volume and we use Eq. \ref{ev3} at a 
%smaller volume which we can assume is in asymptotic 
%regime we obtain
\begin{equation}
P(\hat{Q}_S \geq \hat{q}_S) =\lim_{K\rightarrow \infty} \left[ 1- \left(1- \frac{e^{-\hat{q}_S/2}}{K} \right)^K \right]=1-\exp \left (-e^{-\hat{q}_S/2} \right) .
 \label{sidak}
\end{equation}
Just as equation \ref{pqa} is a generalization of the Bonferroni 
correction, 
equation \ref{sidak} is a generalization of the \v Sid\' ak
correction \cite{sidak} to continuous variables.
%for
%one-tailed tests. 
%For $\hat{q}_S=0$ it gives a
%$p$-value of 0.63, and 
%%In the examples below we will show that this expression applies
%well with simulations 
%%over the entire range of $p$-values, 
%so it is approximately valid 
%%even 
%%in the non-asymptotic regime.
This expression generalizes the $p$-value into the non-asymptotic regime.

%of high $p$-values. 
%%Hence, even though we applied asymptotic approximations throughout the working, by generalizing the \v Sid\' ak correction we have obtained a result that is valid even in the non-asymptotic limit.

\begin{figure}[t]
\hspace*{2.5cm}\includegraphics [width=100mm]{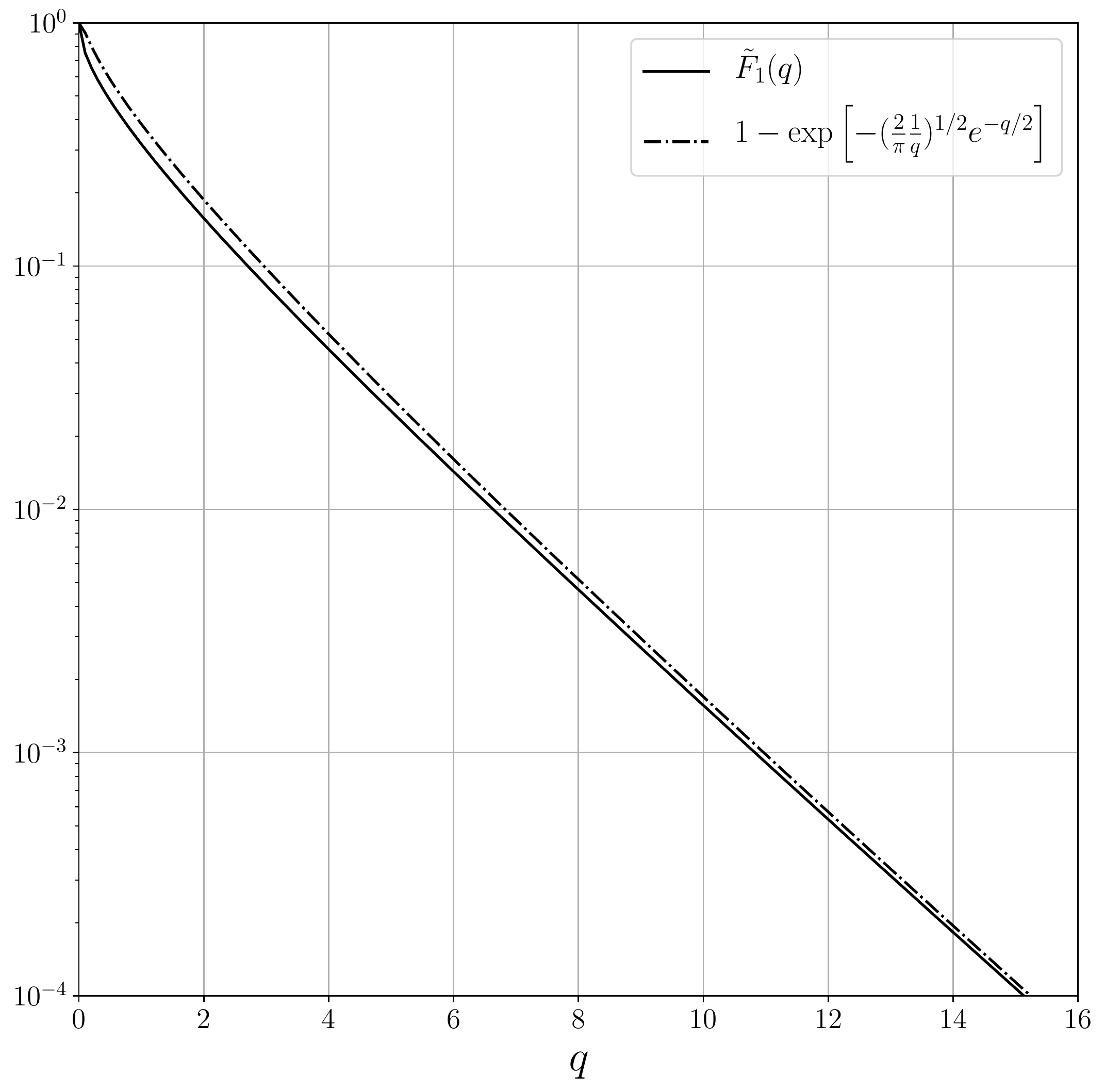}
\caption{Equation \ref{sidak1} is a good approximation to $\tilde{F}_1(q)$ over the entire range of $q$. This suggests that MPS is still accurate in the absence of the look-elsewhere effect for a two-tailed test, even non-asymptotically.}
\label{fig:2tailN1}
\end{figure}

%It is important to recognize that f
For $N\gg 1$ every realization will have a positive peak, hence even the one-tailed 
$p$-value will approach 1 for 
sufficiently low $\hat{q}_L$, which equation \ref{sidak} 
predicts to be for $\hat{q}_S < 0$. 
%Thus, equation \ref{sidak} is expected to agree non-asymptotically in the presence of the look-elsewhere effect.
%
In the absence of the look-elsewhere effect ($N=1$) a 
 one-tailed test should approach a $p$-value of 
 0.5, while equation \ref{sidak} approaches 1 as $\hat{q}_S \rightarrow - \infty$. Thus, the non-asymptotic agreement breaks down for $t=1$ and $N=1$. On the other hand, if $t=2$ and $N=1$, substituting
 $q_S= q_L + \ln 2 \pi q_L-2\ln 2 $ into equation \ref{sidak} gives 
 \begin{equation}
  P(\hat{Q}_L \geq \hat{q}_L)_{N=1,t=2}=1-\exp \left[- \left(\frac{2}{\pi} \frac{1}{\hat{q}_L} \right)^{1/2}e^{-\hat{q}_L/2} \right].
 \label{sidak1}
\end{equation}
The term in the square brackets can be identified as the asymptotic expansion of $\tilde{F}_1(\hat{q}_L)$. We show the non-asymptotic agreement of this equation with $\tilde{F}_1(\hat{q}_L)$, the true two-tailed $p$-value for $N=1$, in figure \ref{fig:2tailN1}. This illustrates the ability of the generalized  \v Sid\' ak correction to produce correct non-asymptotic results, even in the absence of the look-elsewhere effect.
Hence, although we have applied asymptotic approximations throughout the above calculations, we have obtained a result that is valid even in the non-asymptotic limit.
%For $q_L=1, 4, 9$ (1, 2, 3 sigma) it gives $p=0.38, 0.052, 0.029$, compared to the correct two-tailed values of $0.32, 0.045, 0.027$.
% 
Inverting equation \ref{sidak1} gives the significance, or number of sigma, $S$, as
\begin{align}
    S^2 &\approx \hat{q}_S - \ln 2 \pi \hat{q}_S + 2 \ln t, % \\
    %S &= \sqrt{\hat{q}_S} - \frac{1}{2} \frac{\ln 2 \pi \hat{q}_S - 2 \ln 2}{\sqrt{\hat{q}_S}} + \mathcal{O} \left( \frac{1}{\hat{q}_S^{3/2}} \right) 
    %S &= \sqrt{\hat{q}_S} + \mathcal{O} \left( \frac{1}{\sqrt{\hat{q}_S}} \right),
    \label{Nsigma}
\end{align}
with corrections of order $ \mathcal{O} (\hat{q}_S^{-1})$.
In the limit of $\hat{q}_S \rightarrow \infty$, the significance can be interpreted as $\sqrt{\hat{q}_S}$, in an analogous way to $\sqrt{\hat{q}_L}$ in the absence of the look-elsewhere effect. % for a two-tailed test. %We also show the quality of agreement of this formula with the actual value of in figure \ref{fig:2tailN1} 
This motivates the name maximum posterior significance (MPS) as $q_S$ depends on the posterior via the trails factor $N$, and is monotonically related to the significance $S$.

In summary, by considering a frequentist description of the look-elsewhere effect we introduced $\hat{q}_S$ as a natural test statistic to use, such that the asymptotic $p$-value is given by $e^{-\hat{q}_S/2}$. We derived a general expression for the $p$-value which also applies in the non-asymptotic regime, and when there's no look-elsewhere effect. Adopting the prior of equation 
\ref{z1p}, we showed that one can write the
$p$-value in terms of Bayes factor as 
%\begin{align*}
    $\hat{b}^{-1}/\hat{q}_L $. %= e^{-\hat{q}_S}.
%    \label{B-qa}
%\end{align*}
%A corollary of this is that the trials factor is equal to the prior-to-posterior volume ratio of the non-amplitude parameters, which in turn can be used to compute the $p$-value using equation \ref{sidak}.
%This is also the most expensive analysis since 
%it will contain many modes in the posterior, and Bayesian evidence evaluation 
%with multimodal posteriors is a challenging numerical
%problem. 
This intrinsically accounts for the look-elsewhere effect by identifying the trials factor as the prior-to-posterior volume ratio of $\bi{z}_{>1}$ at the MPS mode.
While one can compute the Bayes factor using a variety of methods, we will use the Laplace approximation to evaluate the posterior volume of each mode, as in section \ref{sec:MPM}.
To outline the step-by-step approach:

\vspace{2mm}
\hspace{-0.025\textwidth}
\fcolorbox{black}{cyan!20}{\parbox{0.92\textwidth}{\noindent \textbf{Maximum Posterior Significance (MPS) estimation:}
\begin{enumerate}[nolistsep]
\item Scan over the space of non-amplitude parameters, $\bi{z}_{>1}$, locating peaks in the posterior with any amplitude, $z_1$. Often only the highest few peaks are needed.
\item Compute $q_L$ and the posterior volume, using equation \ref{Vpost}, for each peak.
\item Compute $q_b$ for each peak using equation \ref{qbhat} with the amplitude prior of equation \ref{z1p}.
\item Compute $q_S = q_b + 2 \ln {q_L}$ for each peak. 
\item Find the peak with maximum $q_S$.
\item Compute the (global) $p$-value using equation \ref{sidak} and significance using \ref{Nsigma}.
\end{enumerate}}}
\vspace{2mm}

\subsection{Multiple degrees of freedom}
\label{sec:dofs}

For models with multiple degrees of freedom, the frequentist approach is to apply Wilks' theorem \cite{wilks1938}. This is valid in the asymptotic limit, where, for a two-tailed test, the local $p$-value is given by
\begin{equation}
    P_\nu(\check{Q}_L \geq \check{q}_L)= \tilde{F}_\nu(\check{q}_L) \xrightarrow[]{\check{q}_L \rightarrow \infty}  \frac{1}{\Gamma(\nu/2)} \left( \frac{\check{q}_L}{2} \right)^{\nu/2-1} e^{-\check{q}_L/2},
    \label{pqp}
\end{equation}
for a model with $\nu$ degrees of freedom.
Note that the limit assumes $q \gg \nu$, but for $\nu=2$ it is exact for any $q$.
%$\tilde{F}_s$ is the complementary cumulative distribution function (CCDF) of a chi-squared random variable with $s$ degrees of freedom, which for $s=1$ is the standard normal CCDF multiplied by 2, while for $s=2$ it has a simple analytic expression $\tilde{F}_2(q)=\exp(-q/2)$. %
Wilks' theorem can address the model complexity 
problem of having multiple ($\nu$)
continuous amplitude parameters. 
%To make things concrete 
%it is convenient to introduce a specific model for these 
%amplitude parameters. 
A specific example from particle physics 
is a decay process with $\nu$ decay channels, each with amplitude $A_i$ ($0 \leq i \leq \nu$). In such a case $\max_{\{A_i\}} q_L(\{A_i\}, ...) \sim \tilde{F}_\nu$. 
Wilks' theorem is not sufficiently general:
it fails if the parameters 
are at the edge of their distribution, and it
does not naturally handle the model complexity of the look-elsewhere 
effect, where one scans over a wide range of values for one or more parameters.
Upon introduction of the look-elsewhere effect a frequentist would typically consider single trials distributed as $ \sim \tilde{F}_\nu$, and then use a $\nu$-dependent trials factor \cite{Gross}. Thus in a frequentist approach extra degrees of freedom 
%handled by the Wilks' theorem
and the look-elsewhere effect are treated separately. On the other hand, a Bayesian approach accounts for both in the same way.

To apply the Bayesian methodology, we first reparameterize the model so that there is only a single amplitude parameter by introducing branching ratios $\alpha_i$, such that 
each amplitude parameter is $A_i = \alpha_i z_1$, where $z_1$ 
is the total % proportional to the overall cross-section 
amplitude parameter and $\sum_{i=1}^\nu \alpha_i^2=1$.
To remove the constraint we adopt rotation 
angles: for example, for $\nu=2$ we can work with a phase 
angle $\phi$, such that $\alpha_1=\cos \phi$ and 
$\alpha_2=\sin \phi$. Thus, instead of working with $A_1$ and $A_2$ and considering $\max_{A_1,A_2} q_L(A_1, A_2, ...) \sim \tilde{F}_2$, we consider $\max_{z_1} q_L(z_1, \phi, ...) \sim \tilde{F}_1$ with $\bi{z}_{>1}=(\phi, ...)$. We can then directly apply the MPS prescription for $\nu=1$, as in the previous subsection, by additionally marginalizing over $\phi$ to account for the model complexity with an additional prior-to-posterior volume penalty. 

To be agnostic, one would choose a prior volume for $\phi$ of $V_{\phi, {\rm prior}} = \pi$ (in practice a more complex prior may be appropriate, but it will typically be $\mathcal{O}(1)$). Furthermore, the average error on $\phi$ is typically equal to the relative error on the amplitude, thus $\sigma_{\phi} \approx \sigma_{1}/\mu_1\approx q_L^{-1/2}$. This gives a model complexity correction of 
\begin{align}
    \frac{V_{\phi, {\rm prior}}}{\hat{V}_{\phi, {\rm posterior}}} = \frac{\pi}{\sqrt{2 \pi} \hat{\sigma}_\phi} = \sqrt{\pi} \left( \frac{\hat{q}_L}{2} \right)^{1/2} = \frac{ \tilde{F}_2(\hat{q}_L) }{ \tilde{F}_1(\hat{q}_L) }.
\end{align}
This shows that increasing the model complexity with an extra degree of freedom is accounted for in the Bayesian framework by marginalizing over $\phi$. 
%This is identical to the way in which the Bayesian framework accounts for the look-elsewhere effect.
Thus, the Bayesian answer to an increase in model complexity, whether it be due to including extra degrees of freedom, or looking elsewhere, is identical: 
%reparameterize to a single amplitude
%parameter and compute Bayes factor to determine the $p$-value as in %the $\nu=1$ case. The 
marginalization over the non-amplitude parameters $\bi{z}_{>1}$.
%naturally accounts for the model complexity. 
The $\nu$ dependence of the local 
$p$-value in equation 
\ref{pqp} can be interpreted as a 
Bayesian model complexity penalty: a fixed $p$-value 
corresponds to a larger $\hat{q}_L$ as $\nu$ increases.
Thus, MPS intrinsically generalizes  Wilks’ theorem by relating the trials factor to the prior-to-posterior volume.

%In summary, increasing the model complexity by increasing the number %of degrees of freedom can be accounted for in the Bayesian framework %by marginalizing over rotational angles. 
%Instead of considering the problem as one with 
%$s$ degrees of freedom, reparameterizing to a single amplitude
%parameter $z_1$ and the remaining parameters $z_{>1}$, 
%from which we 
%evaluate the Bayes factor $B_{>1}$. 

\section{Example I: resonance searches}
\label{sec:HEP}

To test the theory of section \ref{sec:bayes_freq} we first consider a resonance search example.
These appear in many different areas of physics, 
including astroparticle and high energy physics.
We consider a search for a new particle whose mass
and cross-section are unknown. The data $\bi{x}$ could correspond to measurements of the invariant mass in the case of collider searches, or the energy flux in astroparticle searches.  The probability density for a single measurement, $x^i$, is given by
\begin{equation}
    p(x^i|f,x_*,\sigma_*)=f p_s(x^i|x_*,\sigma_*)+(1-f)p_b(x^i),
    \label{eqA}
\end{equation}
where $p_s$ and $p_b$ are the normalized signal and background distributions respectively,
%both normalized on the interval $(x_L,x_R)$ with $x_L < x_* < x_R$, 
and $f$ is the fraction of events belonging to the signal.
We assume that the form of the signal and background are known; we take the 
%resonant 
signal to be a normal distribution $p_s(x^i|x_*,\sigma_*) = N(x^i|x_*,\sigma_*)$, and the background to be a power law. % could show more carefully, but whatever. 
Thus the resonance has position $x_*$ and width $\sigma_*$.
Given data $\bi{x}= \{x^i\}_{i=1}^{N_d}$, %, we seek to evaluate the statistical significance of a signal by averaging over many realizations.
%We would also like to determine the cross-section for the resonance, proportional to the number of events that correspond to the resonance events. More specifically, we would like to know the statistical significance of the detection as well as the full posterior on the cross-section, paramaterized by $f$, summarized by its mean, variance, etc.
%
the likelihood is given by the product of the individual probability densities over the data. Using equation \ref{eqA} this gives the likelihood as
\begin{equation}
    p(\bi{x}|f,x_*,\sigma_*)=\prod_{i=1}^{N_d} \left[ f p_s(x^i|x_*,\sigma_*)+(1-f)p_b(x^i) \right].
    \label{HEP_L}
\end{equation}
Note that the Bayesian evidence under the null hypothesis is independent of the parameters, namely
\begin{equation}
    p_0(\bi{x}) \equiv p(\bi{x}|f=0) = \prod_{i=1}^{N_d} p_b(x^i). 
    \label{HEP_L0}
\end{equation}
While the likelihood depends on the number of data $N_d$, quantities such as the $p$-value
%, which are computed by averaging over many realizations, 
will have converged provided $N_d$ is sufficiently large to resolve the resonance. Throughout this section we fix $N_d = 10 V_{x_*,{\rm prior}}/ \sigma_*$ to ensure sufficient convergence. %, and find that anything larger provides identical results. % FIXME: (1) 
We note that more complex models might consider drawing $N_d$ from a Poisson distribution, however this is unnecessary for our proof of concept.

\begin{figure}[t]
\includegraphics [width=\linewidth]{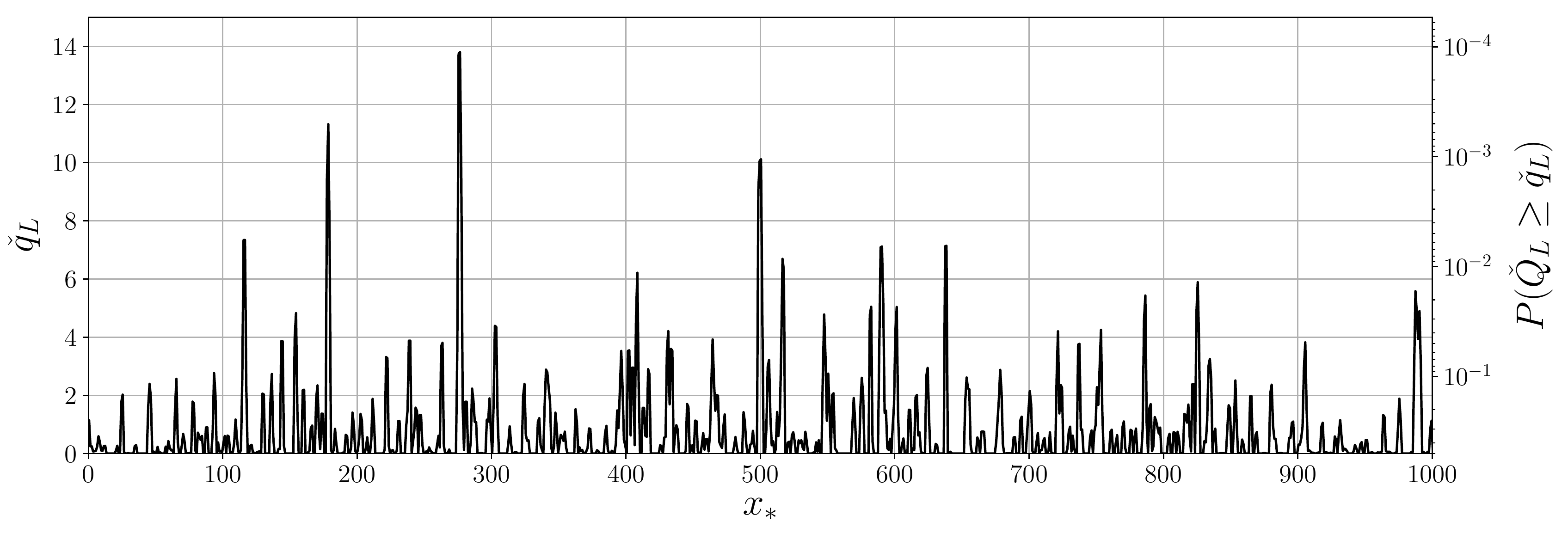}
\caption{The local chi-squared (left axis) and local $p$-value (right axis) for an example data realization with true amplitude $f=5 \times 10^{-3}$, position $x_* = 500$, and width $\sigma_*=0.5$. 
%At each value of $x_*$, the parameter $f$ is set to the value that maximizes the posterior.
While there is a peak with $\check{q}_L \approx 10$ at the correct position, the look-elsewhere effect leads to other, sometimes larger, peaks at random positions.}
\label{fig:particle_peak_locs}
\end{figure}

\begin{figure}[t]
\includegraphics[width=\linewidth]{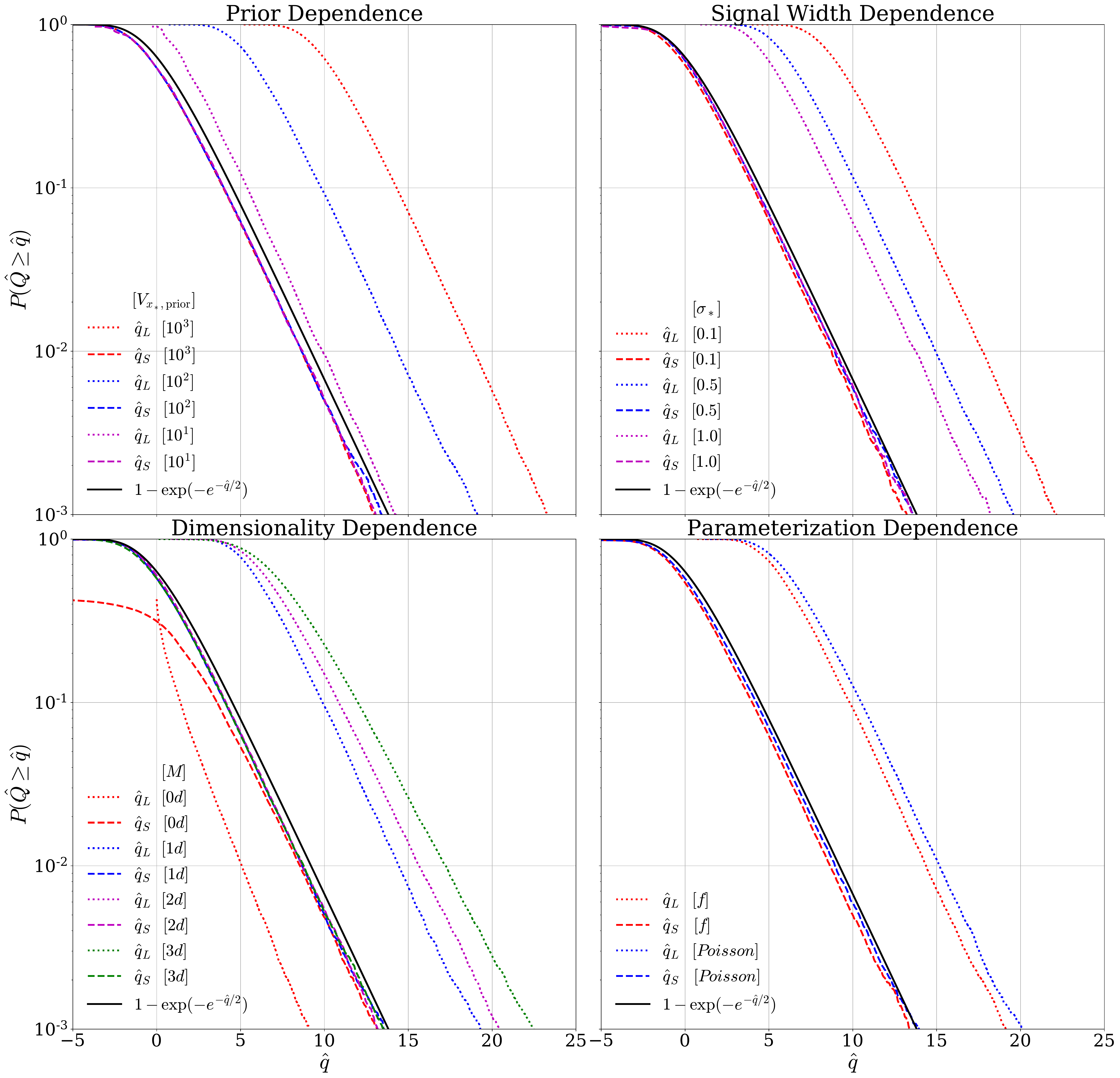}
\caption{
CCDFs of $\hat{q}_L$ (dotted) and $\hat{q}_S$ (dashed), computed using $10^5$ simulations with no signal ($f=0$).
(Top Left) compares three prior volumes: $10^3$ (red), $10^2$ (blue), and $10^1$ (magenta). 
(Top Right) compares different values of signal width $\sigma_*$: 0.1 (red), 0.5 (blue) and 1.0 (magenta).
(Bottom Left) compares the dimensionality of $\bi{x}_*$: $0d$ (red), $1d$ (blue), $2d$ (magenta), and $3d$ (green). (Bottom Right) compares the un-binned $f$-parameterization (red) against a binned Poisson parameterization (blue). 
In all cases the $p$-value of $\hat{q}_L$ has large variation, whereas $\hat{q}_S$ does not. Furthermore, $\hat{q}_S$ closely follows the predictions of equation \ref{sidak} (black).
}
\label{fig:HEP_FPR}
\end{figure}

We first consider a uniform prior on $x_*$, with range $(0,10^3)$, i.e.~a prior volume of $V_{x_*, {\rm prior}} = 10^3$. We do not fit for $\sigma_*$ and fix it to $\sigma_* = 0.5$ a priori, corresponding to the narrow-width approximation.
In this case the posterior is only multimodal in the $x_*$ dimension,
thus to find peaks we split the parameter space along the $x_*$ dimension into narrow 
bins of size $\Delta x_*$ and compute the maximum likelihood 
of equation \ref{HEP_L} within each bin.
%
%: looking at 1 dimensional slices of the posterior, only  will contain multiple peaks.
%, while the $f$ and $\sigma_*$ slices will contain only one. 
Ensuring $\Delta x_*$ is sufficiently small, we determine the location of all peaks in the posterior,  $\bi{\mu}^l$, by comparing adjacent bins. The Hessian at each peak is then computed using finite differencing, and inverted to give $\bi{\Sigma}^l$. Note, in this example we have an analytical form for the likelihood, enabling verification of the numerical computation with analytical results. The value of $q_b$ at each peak is then computed using equation \ref{qbhat}, in turn giving $\hat{q}_S$.

Figure \ref{fig:particle_peak_locs} shows the local chi-squared and local $p$-value as a function of $x_*$ for an example data realization. We use true parameters $f=5 \times 10^{-3}$ and $x_* = 500$. Recall from equation  \ref{pql_local} that the local chi-squared and $p$-value correspond to the values obtained by maximizing over $f$ at fixed $x_*$, i.e.~they correspond to the values obtained without having corrected for the look-elsewhere effect. The local chi-squared $\check{q}_L$ can also be thought of as the projection of $q_L$ onto the $x_*$ axis. It can be seen that although there is a peak with $q_L \approx 10$ at the correct position, there are also multiple spurious peaks throughout the parameter space, with $\hat{q}_L \approx 14$ in this example. %Note, the height of the peak in the correct position is both random and dependent on $f$, we are just showing this figure to illustrate the look-elsewhere effect in action, 
%in the presence of a real anomaly. %Throughout the remainder of this section we will quantify the look-elsewhere effect by averaging over multiple data realizations.
This illustrates the look-elsewhere effect: peaks with a local $p$-value of $\sim 10^{-4}$ are produced by noise, meaning a signal with such a local $p$-value should not be considered as significant as its local $p$-value naively suggests.
%Note that this is just an example realization, in other realizations the peaks would have different heights and positions.

We now consider $10^5$ different data realizations without a signal ($f=0$) to study the distributions of $\hat{q}_L$ and $\hat{q}_S$ under the null hypothesis. The plots in figure \ref{fig:HEP_FPR} show the global $p$-value in terms of $\hat{q}_L$ and $\hat{q}_S$ for a variety of scenarios.
One can think of the vertical axes as corresponding to the false positive rate (FPR) of a hypothesis test using threshold $q$.
%In the context of hypothesis testing, the y-axis is commonly known as the type-1 error, or the false positive rate (FPR): the probability that a signal will be detected when the data does not contain one.

We first compare three different prior volumes on $x_*$, $V_{x_*, {\rm prior}} = 10^3,10^2,10^1$, to show the effectiveness of our method for large and small $N$.
%one with range $(0,10^3)$ (width $10^3$), one with $(450,550)$ (width $10^2$), and the other with $(495,505)$ (width $10^1$). 
The top left plot of figure \ref{fig:HEP_FPR} shows that the $p$-value of $\hat{q}_L$ has a considerable prior volume dependence.  This is the look-elsewhere effect: a larger prior volume leads to a larger trials factor and thus an increased probability of finding a higher maximum likelihood.
On the other hand we see that $\hat{q}_S$ shows no prior dependence and is in good agreement with equation \ref{sidak}, even in the non-asymptotic regime.

We also investigate the variation of the $p$-value with the value of the width of the signal $\sigma_*$. This is shown in the top right plot of figure \ref{fig:HEP_FPR} where we consider $\sigma_* = 0.1, 0.5, 1.0$. Smaller $\sigma_*$ leads to a smaller posterior volume and thus a larger trials factor. Much like the discussion above for prior volume variation, $\hat{q}_L$ has a large $\sigma_*$ dependence, unlike $\hat{q}_S$.

\begin{figure}[t]
\hspace*{2.5cm}\includegraphics [width=100mm]{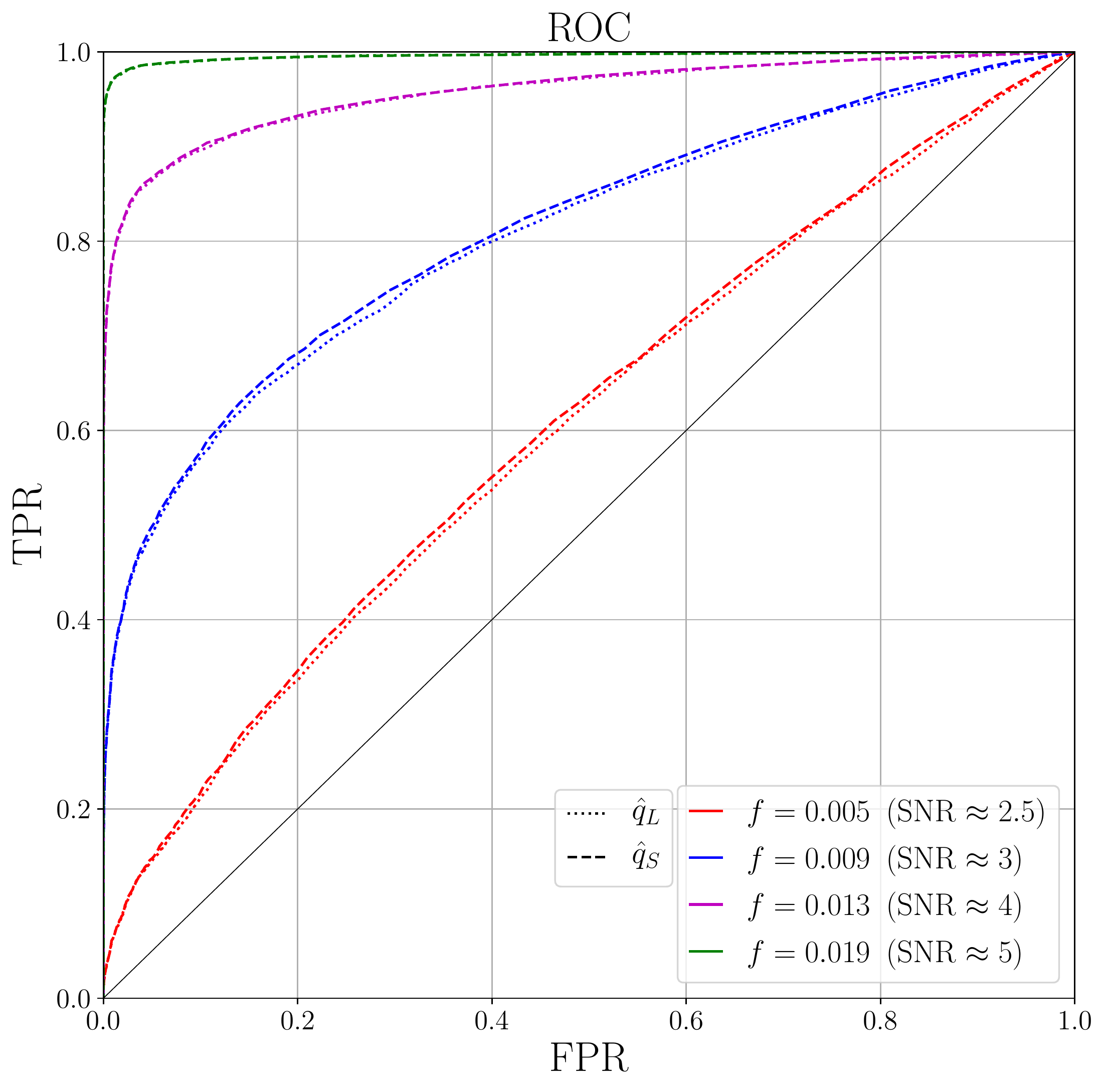}
\caption{ROC curve: comparing the true positive rate (TPR), for a variety of $f$, with the false positive rate (FPR) for $\hat{q}_L$ (dotted) and $\hat{q}_S$ (dashed). 
%A prior volume of 1000 was used. 
%Different $f$ is signified by a different color. 
The signal-to-noise ratio (SNR) corresponds to the average $\sqrt{\hat{q}_L}$ over all data realizations. %All test statistics can be seen to have comparable ROC at a given $f$.
}
\label{fig:ROC}
\end{figure}

Next, we investigate the variation of the $p$-value with the dimensionality of the look-elsewhere effect. To do this we extended the model to consider a signal at vector position $\bi{x}_*$. Each data point now corresponds to a vector $\bi{x}^i$, and we extend the signal and background in a symmetric fashion across each dimension, keeping the total prior volume fixed. Within the context of collider searches, the components of $\bi{x}_*$ might correspond to a collection of invariant mass and jet properties. For astroparticle searches, the multiple dimensions might correspond to different directions in the sky. The bottom left plot of figure \ref{fig:HEP_FPR} shows the variation of the test statistics for dimensionality of 1, 2, and 3, for a constant prior volume of 100. It can be seen that, while the $p$-value of $\hat{q}_L$ is dependent on the dimensionality, the $p$-value of $\hat{q}_S$ is not.
This justifies the naturally arising $(2 \pi)^{M/2}$ prefactor in the posterior volume in equation \ref{Vpost}.
We also plot the $0d$ case, corresponding to only fitting for $A$ with fixed $x_*$. Even though there is no look-elsewhere effect in this case, asymptotic agreement with equation \ref{sidak} is still achieved. This shows our approach is still reliable in the $N \rightarrow 1$ limit, justifying its applicability for arbitrary $N$. As discussed in section \ref{sec:MPS}, non-asymptotic agreement is not expected for a one-tailed test in the absence of the look-elsewhere effect, as the $p$-value tends to 0.5 as $\hat{q}_L \rightarrow 0$; on the other hand, a two-tailed test would give non-asymptotic agreement as shown in figure \ref{fig:2tailN1}. %4

The above discussion concerns an un-binned model, parameterized by the signal fraction $f$. Often in particle physics, one performs a binned analysis with the number of events in each bin modelled as a Poisson distribution \cite{Cowan_2011}.
We find similar results when using this Poisson parameterization, as pictured in the bottom right of figure \ref{fig:HEP_FPR}. The Poisson line agrees with the black line slightly better than the $f$ line does, likely because the Laplace approximation is more accurate in the Poisson case.

When it comes to hypothesis testing, the relation between the true 
positive rate (TPR) and the false positive rate (FPR) determines the predictive power of a test statistic.
In order to compare the relative power of the test statistics we consider an ROC plot for a variety of true $f$ values, shown in figure \ref{fig:ROC}.
%A prior width of $10^3$ was used. 
We also quote the (local) signal-to-noise ratio (SNR), which we define as the average $\sqrt{\hat{q}_L}$ across $10^4$ realizations for the given $f$.
It can be seen that $\hat{q}_S$ 
and $\hat{q}_L$ have approximately equivalent ROC lines, suggesting MAP and MPS have equal predictive power. This is expected as the relation between the test statistics is approximately
monotonic,
%translation
as seen in equation \ref{qa}. Also, it can be seen that the predictive power increases with true $f$ -- as expected a larger true signal is more likely to be correctly detected.

\section{Example II: white noise}
\label{sec:white}

While we could continue the discussion in the context of resonance searches, we now consider a white noise time series example to illustrate the application of MPS to different models. 
This can be thought of as a toy model of a gravitational wave search. In this section we show how MPS handles additional model complexity as theorized in section \ref{sec:dofs}.
We consider a time series $y(x)$ comprising of measurements at $N_d$ times, $\bi{x}= \{x^i\}_{i=1}^{N_d}$, with spacing $x^{i+1} - x^i =1$. In the absence of a signal, each data point $y^i \equiv y(x^i)$ is assumed to be a standard normal random variable, i.e.~we assume white noise. We consider a model with 2 degrees of freedom (dofs), with signal given by
\begin{equation}
    p_s(x|A_1, A_2, x_*, \Delta, \sigma_*) = A_1 N(x | x_*, \sigma_*) + A_2 N(x | x_* + \Delta, \sigma_*)
    \label{ps_white_A12}
\end{equation}
where $A_{1,2} > 0$ are the amplitudes of each dof, $x_*$ and $x_* + \Delta$ are the positions of the dofs, and $\sigma_*$ is the common width.

\begin{figure}[t]
\hspace*{2.0cm}\includegraphics[width=110mm]{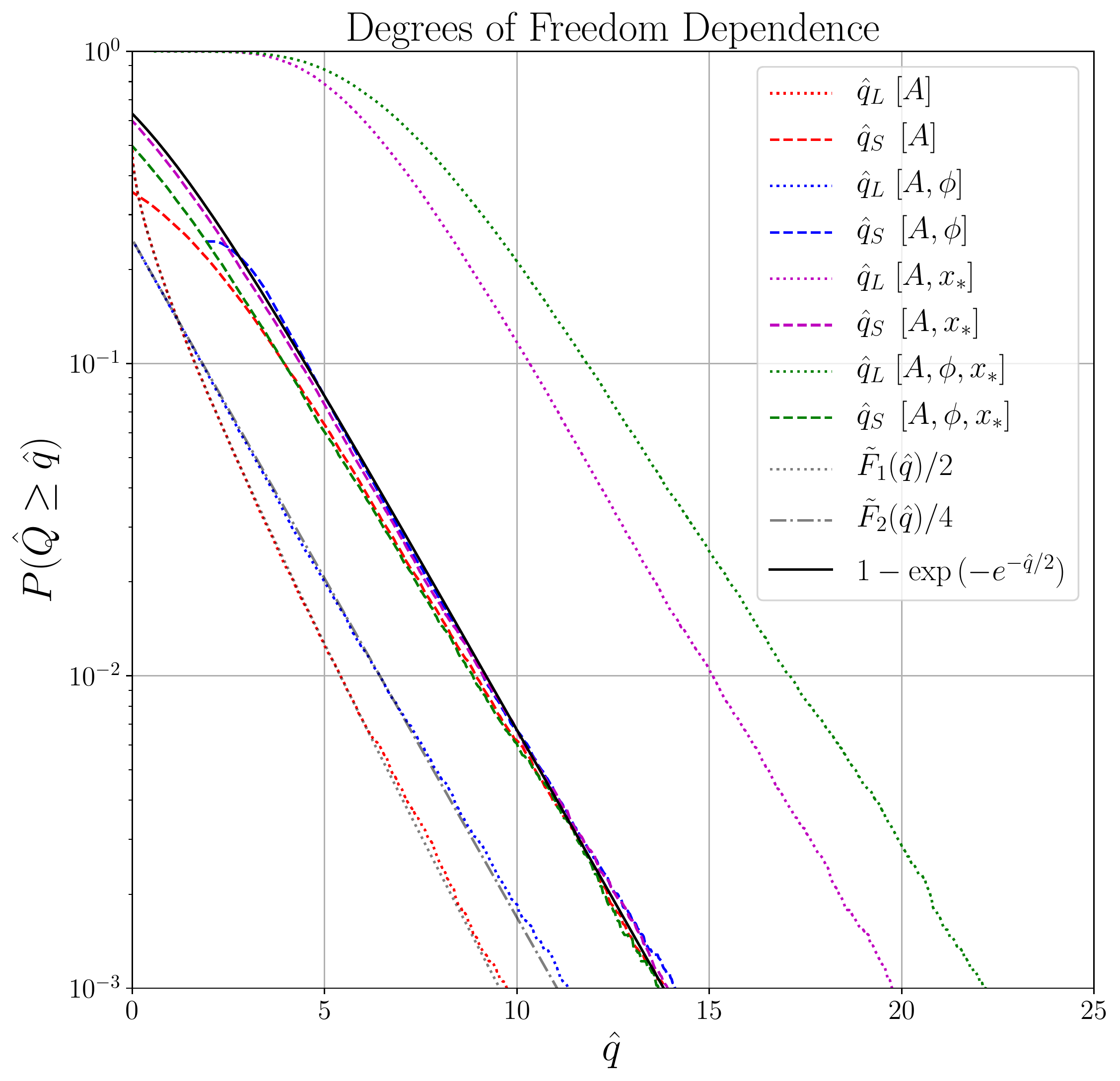}
\caption{
CCDFs of $\hat{q}_L$ and $\hat{q}_S$ averaged over $10^5$ simulations with no signal ($A=0$). The parameters in the square brackets are those being maximized, with other parameters being held fixed (as discussed in the text).
The $p$-value of $\hat{q}_L$ varies depending on the model complexity, whereas $\hat{q}_S$ consistently follows the prediction of equation \ref{sidak} (solid black).
}
\label{fig:dof}
\end{figure}

As motivated in section \ref{sec:dofs}, we reparameterize so that there's a single amplitude parameter, $z_1 = A$, and other parameters describing the properties of the single degree of freedom, $\bi{z}_{>1}$. We thus transform variables using $A_1 = A \cos \phi $ and $A_2 = A \sin \phi$, with $A>0$ and $0 \leq \phi \leq \pi/2$ for a one-tailed test. By substituting the transformations into equation \ref{ps_white_A12}, the signal in the new parameterization is given by
\begin{equation}    
    p_s(x|A, \phi, x_*, \Delta, \sigma_*) = A \left[ \cos \phi N(x | x_*, \sigma_*) + \sin \phi N(x | x_* + \Delta, \sigma_*) \right].
\end{equation}
The corresponding chi-squared difference between the data and the null hypothesis, equal to two times the log-likelihood-ratio, is given by
\begin{align}
    q_L(\bi{x} | A, \phi, x_*, \Delta, \sigma_*) = \sum_{i=1}^{N_d} \left[ y^i - p_s(x^i | A, \phi, x_*, \Delta, \sigma_*) \right]^2 - [y^i]^2.
    \label{white_L}
\end{align}
We consider a uniform prior on $x_*$ with range $(0,100)$, i.e.~a prior volume of $V_{x_*, {\rm prior}} = 100$, and $N_d=100$. We do not fit for $\sigma_*$ or $\Delta$ and fix them to $\sigma_* = 0.5$ and $\Delta=10$.
The application of MPS is identical to the previous section, so we will not repeat the methodology here.

Considering $10^5$ data realizations with no signal, figure \ref{fig:dof} shows how $\hat{q}_L$ and $\hat{q}_S$ are distributed for different levels of model complexity. First we maximize over $A$, while holding all other parameters fixed. In this case $\hat{q}_L \sim \tilde{F}_1(\hat{q}_L)/2$ (red dotted line) as expected for a one-tailed test with one degree of freedom. 
Additionally maximizing over $\phi$ allows for 2 dofs, and gives 
$\hat{q}_L \sim \tilde{F}_2(\hat{q}_L)/4$ (blue dotted line). This is expected because there are 4 permutations of each dof having positive or negative amplitude, and $A_{1,2}>0$ considers 1 of these 4. For both of these cases, $\hat{q}_S$ follows the same asymptotic distribution as predicted by equation \ref{sidak}.  This verifies that the Bayesian picture of marginalizing over $\phi$ successfully reduces a model with 2 dofs to the same scale as 1 dof, in other words Wilks' Theorem has been replaced by marginalizing over $\phi$. There is some discrepancy in the non-asymptotic regime for the maximization over $A$ only (red dashed line), as discussed in section \ref{sec:MPS} for a one-tailed test.

We now introduce the look-elsewhere effect by allowing $x_*$ to vary.
First we maximize over $A$ and $x_*$ for fixed $\phi=0$, as shown by the magenta lines. This is equivalent to a model with 1 dof because $\phi=0$ corresponds to $A_2=0$. We see that the distribution of $\hat{q}_L$ (magenta dotted line) is shifted to the right compared to the red and blue dotted lines due to the look-elsewhere effect. However, the distribution of $\hat{q}_S$ (magenta dashed line) continues to follow the line predicted by equation \ref{sidak}.
Finally, when maximizing over $A$, $\phi$ and $x_*$, i.e.~a model with 2 dofs in the presence of the look-elsewhere effect, $\hat{q}_L$ (green dotted line) is further right-shifted, whereas $\hat{q}_S$ (green dashed line) again agrees with equation \ref{sidak}.
The slight discrepancy in the $A, \phi, x_*$ maximization case is due to using too large a prior volume: there is a slight preference to having two well fitted peaks compared to one very well fitted peak, thus the distribution of $\phi$ is clustered towards $\phi=\pi/4$. Using a more appropriate prior for $\phi$ would improve agreement. 

In summary, while the distribution of $\hat{q}_L$ is highly dependent on the model complexity, via the extra degrees of freedom and look-elsewhere effect, $\hat{q}_S$ has a universal distribution.

\section{Example III: non-Gaussian models of cosmological inflation}
\label{sec:planck}

%\afterpage{
    \begin{figure}[t]
    \includegraphics [width=\linewidth]{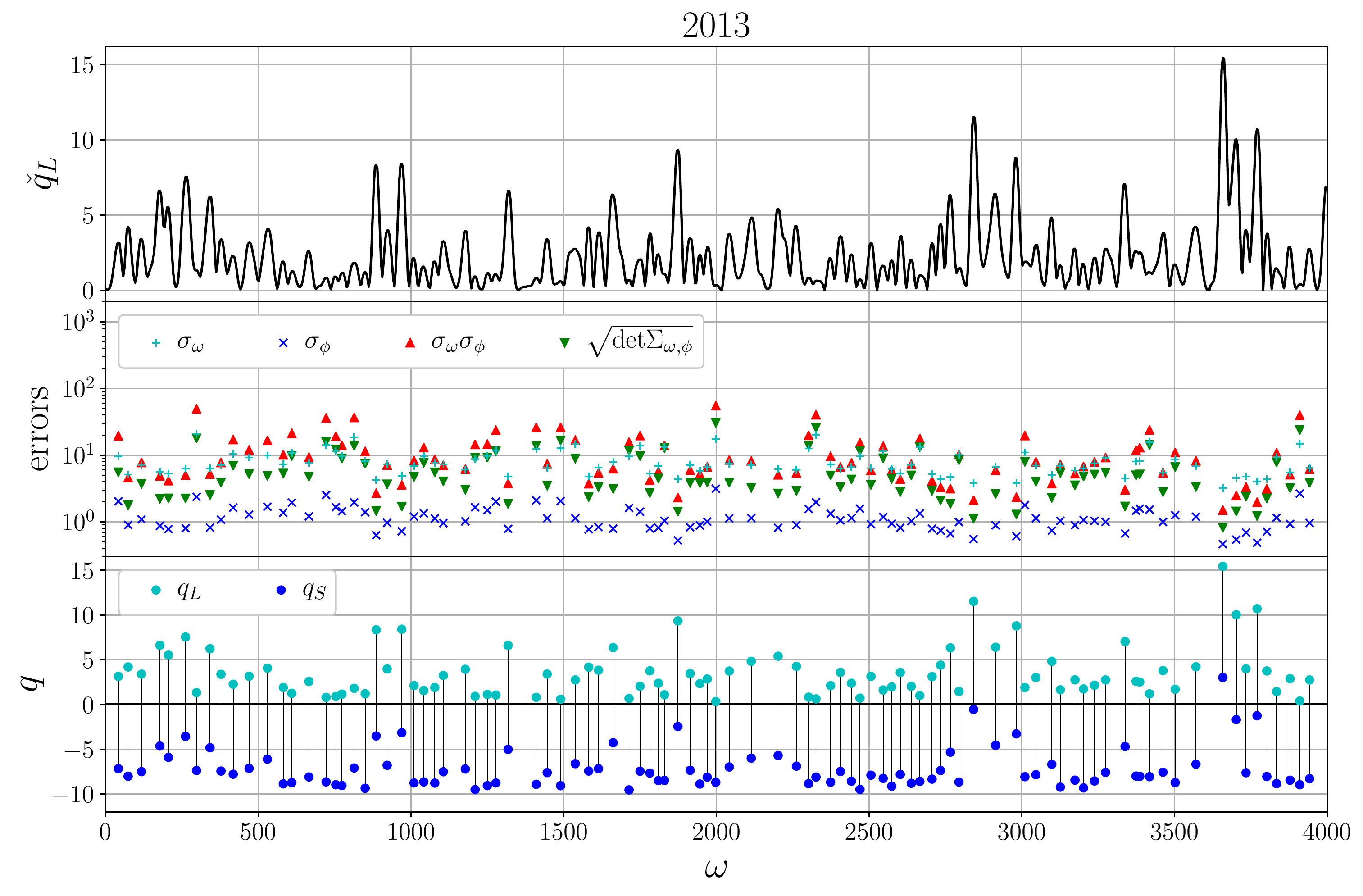}
    \caption{Planck results. Top: Plot of $\check{q}_L$, the projection of $q_L$ onto the $\omega$ axis; this corresponds to $q_L$ evaluated at the $A$ and $\phi$ that maximize $q_L$ at each $\omega$. Middle: The errors obtained for the parameters, as well as a comparison with the determinant of the covariance matrix having removed the amplitude parameter, $\Sigma_{\omega, \phi}$. Bottom: A plot of $q_L$ (blue) and $q_S$ (cyan) for each peak, with the look-elsewhere correction depicted by the vertical black lines.}
    \label{fig:planck}
    \end{figure}
%    \clearpage
%}

There is much interest in detecting non-Gaussian models of inflation via the cosmological power spectrum \cite{Maldacena_2003, Creminelli_2006, Komatsu_2009, Seljak_NG, Planck13_NG, Planck18_NG}. A specific type of such a feature model adds the following oscillatory perturbation to the $\Lambda$CDM power spectrum,
\begin{equation}
    P(k) = P_0(k) [1 + A \sin(2 \omega k + \phi)],
    \label{Pfeature}
\end{equation}
where $P_0(k)$ is the featureless ($\Lambda$CDM) power spectrum and $A$, $\omega$, and $\phi$ are the amplitude, frequency, and phase of the oscillatory perturbation.
%\footnote{We will solely focus on the possibility of a single mode of oscillation, as opposed to a sum of multiple frequencies.} 
Such models are searched for using Planck 2013 data in \cite{Fergusson:2014tza} using the frequentist look-elsewhere analysis technique of \cite{Fergusson:2014hya}. In this section we seek to reproduce the conclusions of these papers using MPS.

Equation \ref{Pfeature} can be written in the form $P(k) = P_0(k) + \Delta P(k)$ with
\begin{equation}
\begin{split}
    \Delta P(k; A, \omega, \phi) &= A P_0(k) [\cos\phi \sin(2 \omega k) + \sin\phi \cos(2 \omega k)] \\
    &\equiv  A  \cos\phi P_s(k; \omega) + A \sin\phi P_c(k; \omega),
    \label{DeltaP}
\end{split}
\end{equation}
where in the last line we explicitly separate terms with $A$ and $\phi$, as only $\omega$ couples to $k$.
Assuming a linear relation, one can write $C_\ell = C_{\ell,0} + \Delta C_\ell$, with
\begin{equation}
    \Delta C_\ell(A, \omega, \phi) = A  \cos\phi C_{\ell,s}(\omega) + A \sin\phi C_{\ell,c}(\omega),
    \label{DeltaCl}
\end{equation}
where $C_{\ell,s}$ and $C_{\ell,c}$ are the angular power spectra corresponding to $P_s$ and $P_c$ respectively.
The Planck Likelihood \cite{Planck13_L} is given by 
\begin{equation}
    -2 \log{L(\hat{C}_\ell | A, \omega, \phi)} = [ \hat{C}_{\ell_1} - C_{\ell_1}(A, \omega, \phi) ] \Delta_{\ell_1 \ell_2} [\hat{C}_{\ell_2} - C_{\ell_2}(A, \omega, \phi)],
\end{equation}
where $\hat{C}_\ell$ are the PCL estimates, and $\Delta_{\ell_1 \ell_2} = \langle \Delta \hat{C}_{\ell_1} \Delta \hat{C}_{\ell_2} \rangle $ is the PCL covariance matrix. In order to compute the likelihood for the null hypothesis, CosmoMC \cite{Lewis_2002} was used to find the best fit values for the cosmological and nuisance parameters. When computing the likelihood for the signal hypothesis, the cosmological parameters were held fixed at these values; while they should really be re-fitted for the signal hypothesis, this is found to have little effect in \cite{Fergusson:2014tza}.
The $C_\ell$ are evaluated using CAMB \cite{Lewis_2000} with a sufficiently high accuracy setting to ensure resolution of the rapid oscillations. To speed up the evaluation of the likelihood over parameter space, $C_{\ell,s}(\omega)$ and $C_{\ell,c}(\omega)$ were computed over a discrete range of $\omega$ between $0$ and $4000$ with step-size $\Delta \omega = 5$, with intermediate values computed via spline interpolation.
A flat prior was chosen for $\omega$ and $\phi$. 
The rest of the analysis is analogous to the previous examples:
we find all the local maxima of the posterior, compute the Hessian using finite differencing, compute the covariance matrix, and use this to find $\hat{q}_S$. Unlike the previous examples, we note that $\omega$ and $\phi$ are correlated, as illustrated in the middle plot of figure \ref{fig:planck}, so it is important to use the determinant of the full covariance matrix and not just its diagonal components. It is also interesting to note that higher peaks have smaller errors.

The results obtained using the CAMspec component of the 2013 Planck likelihood\footnote{One should sum the different components of the likelihood, but this is unnecessary for our proof of concept.} are pictured in 
%the top of 
figure \ref{fig:planck}. The maximum occurs at $\omega\approx3660$ with $\hat{q}_L=15.4$, giving a naive significance of $\sqrt{\hat{q}_L} \approx 4$ sigma.
However, we find that $\hat{q}_S = 3.0$, giving a global $p$-value of $1-\exp(1-e^{-3/2})=0.20$ using equation \ref{sidak}, and significance of $S=1.3$ sigma. Thus the signal is in fact far less significant in light of the look-elsewhere effect. The prescription of  \cite{Fergusson:2014tza} gives a $p$-value of 0.13, which is in reasonable agreement.
Note that our likelihood profile does not match  \cite{Fergusson:2014tza} exactly due to our approximate approach, hence the $p$-value quoted here is the value one would obtain by applying the prescription of  \cite{Fergusson:2014tza} to our likelihood profile.
We applied the same analysis to the 2015 \texttt{plik\_lite} likelihood \cite{Planck15_L} and found a $p$-value of approximately 1, suggesting no evidence for such models of non-Gaussianity.

\section{Conclusions}
\label{sec:conclusion}

This work has employed Bayesian and frequentist thinking to provide a general 
method to account for the look-elsewhere effect.
We started by considering the Bayesian approach, and explained how maximizing the posterior mass, as in MPM, is a more appropriate choice than maximizing the posterior density, as in MAP. Bayesian methodology naturally considers model complexity and the look-elsewhere effect by marginalization, which penalizes the likelihood by the prior-to-posterior volume ratio. Under the Laplace approximation, the 
posterior volume is proportional to 
the determinant of error covariance matrix. 
We then considered the frequentist approach by writing the global $p$-value as the local $p$-value multiplied by the trials factor. By drawing an analogy between the two approaches we identified the trials factor as the prior-to-posterior volume ratio of the parameters being scanned over, in turn generalizing the Bonferroni correction to continuous problems. We introduced $q_S$ and in turn MPS, a hybrid of MPM and MAP, which considers the mode with maximum $q_S$. Finally, we generalized the \v Sid\' ak correction 
to continuous problems, providing a universal way to 
%locate a physical signal in multimodal data and 
assign the global $p$-value in both the asymptotic and non-asymptotic regimes. %, and in the absence of the look-elsewhere effect.

We illustrated the effectiveness of MPS
by considering several examples from 
(astro)particle physics and cosmology, showing it to have equal predictive power to MAP while naturally accounting for the look-elsewhere effect.
MPS effectively shifts the hypothesis testing threshold of the maximum likelihood ratio to a generic scale: while the peak maximum likelihood ratio, or equivalently the best fit chi-squared $\chi^2 = \hat{q}_L$,  depends on the model complexity and extent of the look-elsewhere effect, $\hat{q}_S$ does not. In other words, instead of considering fixed $\hat{q}_L$ thresholds, one should consider
fixed $\hat{q}_S$ thresholds. 

Unlike current methods that rely on performing numerous simulations, MPS accounts for the look-elsewhere effect by using information from the data alone, as one need only compute the likelihood and the posterior volume to evaluate $q_S$. This provides a more efficient way to quantify statistical significance as it does not require expensive simulations. 
In a typical 
situation one would focus on the 
most promising anomalies only, with $\hat{q}_S$ providing a scale that gives good guidance on what false 
positive rate one should expect. Subsequently, one 
would obtain 
additional information to verify the veracity of 
an anomaly when possible. 

For our proof of concept it was sufficient to only consider simple physical examples in this paper, 
but there are many applications where our methods 
can be employed. Examples include searches 
for new particles in astroparticle and particle data, searches for gravitational wave signals in LIGO data,  
searches for exoplanets in transit and radial 
velocity data, as well as many more. In some of these cases the
look-elsewhere penalty can be considerably large, reaching beyond 6 sigma. The problem is very general, as 
almost every search for unknown objects, events, new physics, or other phenomena 
whose existence is unknown,
has to deal with the look-elsewhere
effect. 

The goal of a data analyst 
searching for anomalies is to report the most promising 
anomalies in terms of having a small 
%false positive rate and a high true positive rate, 
$p$-value,
or 
%in terms
%of 
a high Bayes factor. 
%between the alternative and null hypothesis. 
By clarifying the origins of the look-elsewhere effect and 
model complexity penalty for continuous parameters 
we hope to open the way to
refinements in anomaly searches that 
can improve the overall success rate 
of a detection. This should be a common goal of 
any experimental analysis regardless of which school of 
statistics one belongs to.

\acknowledgments

We thank Benjamin Nachman for insightful comments on the manuscript, and Benjamin Wallisch for valuable discussion regarding example III.
This research made use of the Cori supercomputer at the National Energy Research Scientific Computing Center (NERSC), a U.S.~Department of Energy Office of Science User Facility operated under Contract No.~DE-AC02-05CH11231.
This material is based upon work supported by the National Science Foundation under Grant Numbers 1814370 and NSF 1839217, and by NASA under Grant Number 80NSSC18K1274.

\appendix
\section*{Appendices}
\section{Derivation of the CCDF of
\texorpdfstring{\boldmath $\hat{q}_S$}{TEXT}
}
\label{app:qa}

The asymptotic (large $q_L$) CCDF of the global maximum of $q_L$ is for a one-tail test is given in equation \ref{pql_global} as 
\begin{align}
    P_{Q_L}(Q_L \geq q_L) 
    &= N \frac{1}{2} \tilde{F}_1(q_L) \\ 
    &= N \frac{1}{\sqrt{2 \pi q_L}} e^{-q_L/2} + N q_L^{-1/2} e^{-q_L/2}  \mathcal{O}\left(q_L^{-1}\right),
\end{align}
where here we include the leading order correction, and drop hats and take $t=1$ for convenience.
Consider the transformation of variables to $q_S$, defined by
\begin{align}
    q_S \equiv g(q_L) \equiv q_L - 2 \ln N + \ln 2 \pi q_L.
\end{align}
It can be shown that the inverse of $g$ is given by
\begin{align}
    q_L = g^{-1}(q_S) 
    &= W_0\left( \frac{N^2 e^{q_S}}{2 \pi} \right) \\
    &= q_S + \ln \frac{N^2}{2 \pi} - \ln \left( q_S + \ln \frac{N^2}{2 \pi}  \right) + \mathcal{O}\left( \frac{L_2}{L_1} \right),
\end{align}
where $W_0(z)$ is the principal branch of the Lambert $W$ function. The asymptotic expansion has been performed in the final line, with the shorthand $L_i \equiv \ln^i \frac{N^2 e^{q_S}}{2 \pi} $. Assuming $N$ is constant to study the limiting behaviour, the CCDF of $q_S$ is thus
\begin{align}
    P_{Q_a} &\left( Q_a \geq q_S \right)
    = P_{Q_L}\left[Q_L \geq g^{-1}(q_S)\right] \\
    &= e^{- q_S /2} e^{-\mathcal{O}\left( L_2 / L_1 \right)} \left( 1 - \frac{\ln \left( q_S + \ln \frac{N^2}{2 \pi}  \right) + \mathcal{O}\left( \frac{L_2}{L_1} \right)}{q_S + \ln \frac{N^2}{2 \pi}} \right)^{-1/2} + \mathcal{O}\left( \frac{e^{-q_S/2}}{q_S + \ln \frac{N^2}{2 \pi}} \right) \\
    &\rightarrow e^{- q_S /2},
\end{align}
where the limit corresponds to either $N \rightarrow \infty$ or $q_S \rightarrow \infty$. This means the result still applies asymptotically in the absence of the look-elsewhere effect ($N=1$).

\bibliography{refs}

\providecommand{\href}[2]{#2}\begingroup\raggedright\begin{thebibliography}{10}

\bibitem{Miller_1981}
R.~G. Miller, \emph{Simultaneous Statistical Inference}. Springer New York,
  1981,
  \href{https://doi.org/10.1007/978-1-4613-8122-8}{10.1007/978-1-4613-8122-8}.

\bibitem{mult_hyp_test}
J.~P. Shaffer, \emph{Multiple hypothesis testing},
  \href{https://doi.org/10.1146/annurev.ps.46.020195.003021}{\emph{Annual
  Review of Psychology} {\bfseries 46} (1995) 561}
  [\href{https://arxiv.org/abs/https://doi.org/10.1146/annurev.ps.46.020195.003021}{{\ttfamily
  https://doi.org/10.1146/annurev.ps.46.020195.003021}}].

\bibitem{Aad:2012tfa}
{\scshape ATLAS} collaboration, G.~Aad et~al., \emph{{Observation of a new
  particle in the search for the Standard Model Higgs boson with the ATLAS
  detector at the LHC}},
  \href{https://doi.org/10.1016/j.physletb.2012.08.020}{\emph{Phys. Lett.}
  {\bfseries B716} (2012) 1} [\href{https://arxiv.org/abs/1207.7214}{{\ttfamily
  1207.7214}}].

\bibitem{Chatrchyan:2012xdj}
{\scshape CMS} collaboration, S.~Chatrchyan et~al., \emph{{Observation of a New
  Boson at a Mass of 125 GeV with the CMS Experiment at the LHC}},
  \href{https://doi.org/10.1016/j.physletb.2012.08.021}{\emph{Phys. Lett.}
  {\bfseries B716} (2012) 30}
  [\href{https://arxiv.org/abs/1207.7235}{{\ttfamily 1207.7235}}].

\bibitem{Anderson_2016}
B.~Anderson, S.~Zimmer, J.~Conrad, M.~Gustafsson, M.~Sánchez-Conde and
  R.~Caputo, \emph{Search for gamma-ray lines towards galaxy clusters with the
  fermi-lat},
  \href{https://doi.org/10.1088/1475-7516/2016/02/026}{\emph{Journal of
  Cosmology and Astroparticle Physics} {\bfseries 2016} (2016) 026–026}.

\bibitem{Reinert_2018}
A.~Reinert and M.~W. Winkler, \emph{A precision search for {WIMPs} with charged
  cosmic rays},
  \href{https://doi.org/10.1088/1475-7516/2018/01/055}{\emph{Journal of
  Cosmology and Astroparticle Physics} {\bfseries 2018} (2018) 055}.

\bibitem{10.1093/pasj/psv081}
N.~Sekiya, N.~Y. Yamasaki and K.~Mitsuda, \emph{{A search for a keV signature
  of radiatively decaying dark matter with Suzaku XIS observations of the X-ray
  diffuse background}},
  \href{https://doi.org/10.1093/pasj/psv081}{\emph{Publications of the
  Astronomical Society of Japan} {\bfseries 68} (2015) }
  [\href{https://arxiv.org/abs/https://academic.oup.com/pasj/article-pdf/68/SP1/S31/7971976/psv081.pdf}{{\ttfamily
  https://academic.oup.com/pasj/article-pdf/68/SP1/S31/7971976/psv081.pdf}}].

\bibitem{Aartsen_2014}
M.~Aartsen, M.~Ackermann, J.~Adams, J.~Aguilar, M.~Ahlers, M.~Ahrens et~al.,
  \emph{Observation of high-energy astrophysical neutrinos in three years of
  icecube data},
  \href{https://doi.org/10.1103/physrevlett.113.101101}{\emph{Physical Review
  Letters} {\bfseries 113} (2014) }.

\bibitem{Emig_2015}
K.~Emig, C.~Lunardini and R.~Windhorst, \emph{Do high energy astrophysical
  neutrinos trace star formation?},
  \href{https://doi.org/10.1088/1475-7516/2015/12/029}{\emph{Journal of
  Cosmology and Astroparticle Physics} {\bfseries 2015} (2015) 029–029}.

\bibitem{cannon2015likelihoodratio}
K.~Cannon, C.~Hanna and J.~Peoples, \emph{Likelihood-ratio ranking statistic
  for compact binary coalescence candidates with rate estimation},  2015.

\bibitem{Abbott_2016}
B.~Abbott, R.~Abbott, T.~Abbott, M.~Abernathy, F.~Acernese, K.~Ackley et~al.,
  \emph{Gw150914: First results from the search for binary black hole
  coalescence with advanced ligo},
  \href{https://doi.org/10.1103/physrevd.93.122003}{\emph{Physical Review D}
  {\bfseries 93} (2016) }.

\bibitem{Messick_2017}
C.~Messick, K.~Blackburn, P.~Brady, P.~Brockill, K.~Cannon, R.~Cariou et~al.,
  \emph{Analysis framework for the prompt discovery of compact binary mergers
  in gravitational-wave data},
  \href{https://doi.org/10.1103/physrevd.95.042001}{\emph{Physical Review D}
  {\bfseries 95} (2017) }.

\bibitem{Fergusson:2014hya}
J.~R. Fergusson, H.~F. Gruetjen, E.~P.~S. Shellard and M.~Liguori,
  \emph{{Combining power spectrum and bispectrum measurements to detect
  oscillatory features}},
  \href{https://doi.org/10.1103/PhysRevD.91.023502}{\emph{Phys. Rev.}
  {\bfseries D91} (2015) 023502}
  [\href{https://arxiv.org/abs/1410.5114}{{\ttfamily 1410.5114}}].

\bibitem{Fergusson:2014tza}
J.~R. Fergusson, H.~F. Gruetjen, E.~P.~S. Shellard and B.~Wallisch,
  \emph{{Polyspectra searches for sharp oscillatory features in cosmic
  microwave sky data}},
  \href{https://doi.org/10.1103/PhysRevD.91.123506}{\emph{Phys. Rev.}
  {\bfseries D91} (2015) 123506}
  [\href{https://arxiv.org/abs/1412.6152}{{\ttfamily 1412.6152}}].

\bibitem{Hunt_2015}
P.~Hunt and S.~Sarkar, \emph{Search for features in the spectrum of primordial
  perturbations using planck and other datasets},
  \href{https://doi.org/10.1088/1475-7516/2015/12/052}{\emph{Journal of
  Cosmology and Astroparticle Physics} {\bfseries 2015} (2015) 052–052}.

\bibitem{robnik2019kepler}
J.~Robnik and U.~Seljak, \emph{Kepler data analysis: non-gaussian noise and
  fourier gaussian process analysis of star variability},  2019.

\bibitem{Genes1}
P.~I.~W. de~Bakker, R.~Yelensky, I.~Pe'er, S.~B. Gabriel, M.~J. Daly and
  D.~Altshuler, \emph{Efficiency and power in genetic association studies},
  \href{https://doi.org/10.1038/ng1669}{\emph{Nature Genetics} {\bfseries 37}
  (2005) 1217}.

\bibitem{Genes2}
J.~D. Storey and R.~Tibshirani, \emph{Statistical significance for genomewide
  studies}, \href{https://doi.org/10.1073/pnas.1530509100}{\emph{Proceedings of
  the National Academy of Sciences of the United States of America} {\bfseries
  100} (2003) 9440}.

\bibitem{drugs}
M.~Proschan and M.~Waclawiw, \emph{Practical guidelines for multiplicity
  adjustment in clinical trials},
  \href{https://doi.org/10.1016/S0197-2456(00)00106-9}{\emph{Controlled
  clinical trials} {\bfseries 21} (2001) 527}.

\bibitem{BibleCode2_mckay1999}
B.~McKay, D.~Bar-Natan, M.~Bar-Hillel and G.~Kalai, \emph{Solving the bible
  code puzzle}, \href{https://doi.org/10.1214/ss/1009212243}{\emph{Statist.
  Sci.} {\bfseries 14} (1999) 150}.

\bibitem{bonferroni}
C.~E. Bonferroni, \emph{Teoria statistica delle classi e calcolo delle
  probabilità}, {\emph{Pubblicazioni del R Istituto Superiore di Scienze
  Economiche e Commerciali di Firenze} (1936) }.

\bibitem{sidak}
Z.~Šidák, \emph{Rectangular confidence regions for the means of multivariate
  normal distributions},
  \href{https://doi.org/10.1080/01621459.1967.10482935}{\emph{Journal of the
  American Statistical Association} {\bfseries 62} (1967) 626}
  [\href{https://arxiv.org/abs/https://doi.org/10.1080/01621459.1967.10482935}{{\ttfamily
  https://doi.org/10.1080/01621459.1967.10482935}}].

\bibitem{Gross}
E.~Gross and O.~Vitells, \emph{Trial factors for the look elsewhere effect in
  high energy physics},
  \href{https://doi.org/10.1140/epjc/s10052-010-1470-8}{\emph{The European
  Physical Journal C} {\bfseries 70} (2010) 525–530}.

\bibitem{Planck13_L}
{Planck Collaboration}, P.~A.~R. {Ade}, N.~{Aghanim}, C.~{Armitage-Caplan},
  M.~{Arnaud}, M.~{Ashdown} et~al., \emph{{Planck 2013 results. XV. CMB power
  spectra and likelihood}},
  \href{https://doi.org/10.1051/0004-6361/201321573}{\emph{Astronomy \&
  Astrophysics} {\bfseries 571} (2014) A15}
  [\href{https://arxiv.org/abs/1303.5075}{{\ttfamily 1303.5075}}].

\bibitem{Chen2000}
M.-H. Chen, Q.-M. Shao and J.~G. Ibrahim, \emph{Estimating Ratios of
  Normalizing Constants}, pp.~124--190.
\newblock Springer New York, New York, NY, 2000.
\newblock 10.1007/978-1-4612-1276-8\_5.

\bibitem{HeSeljak}
H.~Jia and U.~Seljak, \emph{Normalizing constant estimation with gaussianized
  bridge sampling},  2019.

\bibitem{SeljakYu2019}
U.~Seljak and B.~Yu, \emph{Posterior inference unchained with $\rm{EL_2O}$},
  2019.

\bibitem{MacKay2003}
D.~J.~C. MacKay, \emph{Information Theory, Inference, and Learning Algorithms}.
  Copyright Cambridge University Press, 2003.

\bibitem{Lindley}
D.~V. Lindley, \emph{A statistical paradox}, {\emph{Biometrika} {\bfseries 44}
  (1957) 187}.

\bibitem{Cousins_2014}
R.~D. Cousins, \emph{The jeffreys–lindley paradox and discovery criteria in
  high energy physics},
  \href{https://doi.org/10.1007/s11229-014-0525-z}{\emph{Synthese} {\bfseries
  194} (2014) 395–432}.

\bibitem{Nesseris_2013}
S.~Nesseris and J.~García-Bellido, \emph{Is the jeffreys’ scale a reliable
  tool for bayesian model comparison in cosmology?},
  \href{https://doi.org/10.1088/1475-7516/2013/08/036}{\emph{Journal of
  Cosmology and Astroparticle Physics} {\bfseries 2013} (2013) 036–036}.

\bibitem{wilks1938}
S.~S. Wilks, \emph{The large-sample distribution of the likelihood ratio for
  testing composite hypotheses},
  \href{https://doi.org/10.1214/aoms/1177732360}{\emph{Ann. Math. Statist.}
  {\bfseries 9} (1938) 60}.

\bibitem{Davies}
R.~B. Davies, \emph{Hypothesis testing when a nuisance parameter is present
  only under the alternative},
  \href{https://doi.org/10.2307/2335690}{\emph{Biometrika} {\bfseries 64}
  (1977) 247}.

\bibitem{Davies2}
R.~B. Davies, \emph{{Hypothesis testing when a nuisance parameter is present
  only under the alternative}},
  \href{https://doi.org/10.1093/biomet/74.1.33}{\emph{Biometrika} {\bfseries
  74} (1987) 33}.

\bibitem{Cowan_2011}
G.~Cowan, K.~Cranmer, E.~Gross and O.~Vitells, \emph{Asymptotic formulae for
  likelihood-based tests of new physics},
  \href{https://doi.org/10.1140/epjc/s10052-011-1554-0}{\emph{The European
  Physical Journal C} {\bfseries 71} (2011) }.

\bibitem{Maldacena_2003}
J.~Maldacena, \emph{Non-gaussian features of primordial fluctuations in single
  field inflationary models},
  \href{https://doi.org/10.1088/1126-6708/2003/05/013}{\emph{Journal of High
  Energy Physics} {\bfseries 2003} (2003) 013–013}.

\bibitem{Creminelli_2006}
P.~Creminelli, A.~Nicolis, L.~Senatore, M.~Tegmark and M.~Zaldarriaga,
  \emph{Limits on non-gaussianities from wmap data},
  \href{https://doi.org/10.1088/1475-7516/2006/05/004}{\emph{Journal of
  Cosmology and Astroparticle Physics} {\bfseries 2006} (2006) 004–004}.

\bibitem{Komatsu_2009}
E.~Komatsu, J.~Dunkley, M.~R. Nolta, C.~L. Bennett, B.~Gold, G.~Hinshaw et~al.,
  \emph{Five-year wilkinson microwave anisotropy probe observations:
  Cosmological interpretation},
  \href{https://doi.org/10.1088/0067-0049/180/2/330}{\emph{The Astrophysical
  Journal Supplement Series} {\bfseries 180} (2009) 330–376}.

\bibitem{Seljak_NG}
U.~Seljak, \emph{Extracting primordial non-gaussianity without cosmic
  variance}, \href{https://doi.org/10.1103/PhysRevLett.102.021302}{\emph{Phys.
  Rev. Lett.} {\bfseries 102} (2009) 021302}.

\bibitem{Planck13_NG}
{Planck Collaboration}, P.~A.~R. {Ade}, N.~{Aghanim}, C.~{Armitage-Caplan},
  M.~{Arnaud}, M.~{Ashdown} et~al., \emph{{Planck 2013 results. XXIV.
  Constraints on primordial non-Gaussianity}},
  \href{https://doi.org/10.1051/0004-6361/201321554}{\emph{Astronomy \&
  Astrophysics} {\bfseries 571} (2014) A24}
  [\href{https://arxiv.org/abs/1303.5084}{{\ttfamily 1303.5084}}].

\bibitem{Planck18_NG}
P.~Collaboration, Y.~Akrami, F.~Arroja, M.~Ashdown, J.~Aumont, C.~Baccigalupi
  et~al., \emph{Planck 2018 results. ix. constraints on primordial
  non-gaussianity},  2019.

\bibitem{Lewis_2002}
A.~Lewis and S.~Bridle, \emph{Cosmological parameters from cmb and other data:
  A monte carlo approach},
  \href{https://doi.org/10.1103/physrevd.66.103511}{\emph{Physical Review D}
  {\bfseries 66} (2002) }.

\bibitem{Lewis_2000}
A.~Lewis, A.~Challinor and A.~Lasenby, \emph{Efficient computation of cosmic
  microwave background anisotropies in closed friedmann-robertson-walker
  models}, \href{https://doi.org/10.1086/309179}{\emph{The Astrophysical
  Journal} {\bfseries 538} (2000) 473}.

\bibitem{Planck15_L}
{\scshape Planck} collaboration, N.~Aghanim et~al., \emph{{Planck 2015 results.
  XI. CMB power spectra, likelihoods, and robustness of parameters}},
  \href{https://doi.org/10.1051/0004-6361/201526926}{\emph{Astron. Astrophys.}
  {\bfseries 594} (2016) A11}
  [\href{https://arxiv.org/abs/1507.02704}{{\ttfamily 1507.02704}}].

\end{thebibliography}\endgroup
\bibliographystyle{JHEP}

\end{document}